\documentclass[prb,superscriptaddress]{revtex4-1}

\usepackage{graphicx}
\usepackage{amsmath}
\usepackage{amssymb}
\usepackage{float}
\usepackage{subfig}

\begin{document}
\title{Indications of spin polarized transport in Ba$_2$FeMoO$_6$ thin films}

\author{S.~Granville} \email{simon.granville@vuw.ac.nz}
  \affiliation{Robinson Research Institute, Victoria University of Wellington, PO Box 600, Wellington, New Zealand}
  \affiliation{The MacDiarmid Institute for Advanced Materials and Nanotechnology, New Zealand}
\author{I.~L.~Farrell} 
  \affiliation{Department of Physics and Astronomy, University of Canterbury, Private Bag 4800, Christchurch, New Zealand}
\author{A.~R.~Hyndman}
  \affiliation{Industrial Research Ltd., Lower Hutt, New Zealand}
\author{D.~M.~McCann}
  \affiliation{Industrial Research Ltd., Lower Hutt, New Zealand}
\author{R.~J.~Reeves}
  \affiliation{The MacDiarmid Institute for Advanced Materials and Nanotechnology, New Zealand}
  \affiliation{Department of Physics and Astronomy, University of Canterbury, Private Bag 4800, Christchurch, New Zealand}
\author{G.~V.~M.~Williams}
 \affiliation{The MacDiarmid Institute for Advanced Materials and Nanotechnology, New Zealand}
 \affiliation{School of Chemical and Physical Sciences, Victoria University of Wellington, PO Box 600, Wellington, New Zealand}

\begin{abstract}
We have investigated the magnetic and magnetotransport properties of  Ba$_2$FeMoO$_6$ thin films produced by pulsed laser deposition from optimized bulk material.  The films are comprised of grains of crystalline  Ba$_2$FeMoO$_6$ with a disordered grain boundary region that lowers the net saturation magnetization of the film and prevents full magnetic alignment below a Curie temperature $T$\textsubscript{C}$\sim$305~K.  Magnetotransport measurements point to the Ba$_2$FeMoO$_6$ grains retaining the high spin polarization of a half-metal up to $T$\textsubscript{C}, while the grain boundaries greatly reduce the spin polarization of the intergrain electrical current due to spin-flip scattering. Our results show that a strong spin polarization of the electronic charge carriers is present even in  Ba$_2$FeMoO$_6$ films that do not show the ideal bulk magnetic character.  

\end{abstract}

\date{\today}
\maketitle

\section{Introduction}

The double perovskite oxide compounds A$_2$BBO$_6$ have excited interest due to the expectation that a number of them should be exceptional materials for enabling new spin-based devices for electronics and computing~\cite{kobayashi_1998,bibes_2007,hirohata_2015}.  In particular, a number of the double perovskites are predicted to be half-metallic with a 100\% spin polarization of their charge carriers~\cite{kobayashi_1998,borges_1999,szotek_electronic_2003}.  The most extensively studied double perovskite oxide is Sr$_2$FeMoO$_6$, owing to its high Curie temperature ($T$\textsubscript{C}) in the bulk of 426-440~K~\cite{borges_1999,itoh_1996}. A significant effort has gone into optimizing the preparation of this compound in bulk ceramic form~\cite{serrate_double_2007}, with promising experimental results demonstrated, such as magnetoresistances that remain as large as 10\% at room temperature~\cite{
retuerto_2004,huang_2005,aldica_2007}.  However, producing thin films of Sr$_2$FeMoO$_6$ with bulk-like properties has proven to be difficult, and more often than not the saturation magnetization and $T$\textsubscript{C} of films have been lower than for the bulk~\cite{sanchez_structural_2004,di_trolio_microstructure_2006,wang_double-perovskite_2006,song_double-perovskite_2008,hauser_unlocking_2011}. In addition, the magnetoresistance of thin films has typically not exceeded 5\%~\cite{shinde_2003,deniz_2017}, lower than the magnetoresistances of bulk material, which reach as high as 15\%~\cite{aldica_2007}. Recently however, a room temperature magnetoresistance of 12\% has been achieved in a thin film deposited on Si~\cite{kumar_polycrystalline_2014}, and films have also been incorporated into magnetic tunnel junction structures~\cite{bibes_tunnel_2003,fix_diode_2006,kumar_room_2014}.  

The next-most investigated member of the family is Ba$_2$FeMoO$_6$ (BFMO) , also predicted to be a half-metal~\cite{szotek_electronic_2003,kang_electronic_2002,mohamed_musa_saad_first-principles_2012}, with a similar magnetic moment to Sr$_2$FeMoO$_6$, and a $T$\textsubscript{C} higher than room temperature ($T$\textsubscript{C}$\approx$330-345~K)~\cite{maignan_large_1999,han_effect_2001,kim_neutron_2002,pandey_magnetic_2009}.  BFMO has also been extensively investigated in bulk ceramic form and several groups have produced material with very near the theoretical maximum saturation magnetization~\cite{maignan_large_1999,hemery_effect_2007,pandey_magnetic_2009,yang_enhancement_2003}.  However, only a few studies of the thin film form of BFMO exist.  Thin polycrystalline films prepared by ion beam sputtering from bulk targets followed by high temperature annealing have achieved saturation magnetization of 3.27~$\mu$\textsubscript{B} per formula unit (f.u.)~\cite{mccann_large_2014}.  Crystalline epitaxial films have been prepared by pulsed laser deposition (PLD) at high temperatures, with a lower saturation magnetization than achieved in the bulk form. These PLD films had a $T$\textsubscript{C} of 250~K when grown in vacuum~\cite{kim_magnetic_2012}, which increased to above 300~K when grown in an Ar/O$_2$ partial pressure~\cite{kim_effects_2015}.  However, the films grown in an O$_2$ environment had a reduced saturation magnetization and showed signs of phase segregation in structural measurements.  

To further advance the development of BFMO as a candidate material for spin-device technology, we have deposited thin films from bulk targets prepared with optimum properties for BFMO~\cite{hemery_effect_2007,mccann_large_2014,stephen_magnetic_2013}.  The films were deposited by pulsed laser deposition at the elevated temperatures required for epitaxial growth on lattice-matched substrates~\cite{kim_magnetic_2012}.  We have characterized the magnetic and magnetotransport properties of these films prepared from optimized targets in order to investigate their potential use as a source of spin polarized charge carriers. 

\section{Growth and Methods}

\begin{figure}
\includegraphics[width=9cm]{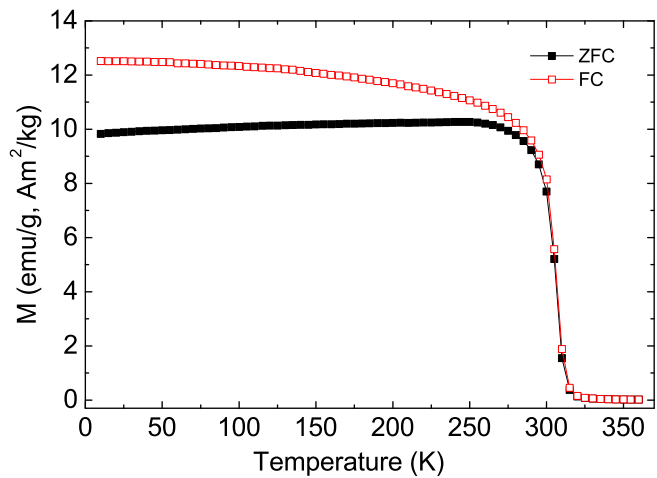}
\caption{(color online) Magnetization vs temperature of the optimized Ba$_2$FeMoO$_6$ pellet used as a target for the growth of thin films via pulsed laser deposition, measured in 100~Oe ($\sim$8000~Am$^{-1}$)~field.  $T$\textsubscript{C} is approximately 315~K.
\label{MT}}
\vspace{-0.5 cm}\end{figure}

We have first produced BFMO in bulk pellet form by solid state reaction, grinding and sintering powders containing the constituent elements.  The preparation details for the bulk pellets have been reported elsewhere~\cite{mccann_large_2014,farrell_2017}.  Our optimized pellet used as the target for film growth has a $T$\textsubscript{C}=315$\pm$2~K as shown in Figure~\ref{MT}, within the range of 310-330~K usually reported for bulk BFMO~\cite{maignan_large_1999,han_effect_2001,kim_neutron_2002,pandey_magnetic_2009,kim_sintering_2011}.  The saturation magnetization $M$\textsubscript{S} of the target at 10~K is 3.72~$\mu$\textsubscript{B}, close to the theoretical maximum of 4~$\mu$\textsubscript{B}~\cite{borges_1999}. Using the equation $M$\textsubscript{S}=$(4-8x)\mu$\textsubscript{B} where $x$ is the anti-site disorder (ASD)~\cite{ogale_octahedral_1999,balcells_cationic_2001,niebieskikwiat_antisite_2004}, $x$ for the target is calculated as 3.4$\%$, among the lowest values reported for BFMO to date~\cite{hemery_effect_2007,pandey_magnetic_2009,han_magnetic_2003}.

Thin films with thicknesses $\sim$600-700~nm have been grown from the optimized target by pulsed laser deposition onto SrTiO$_3$ (001) substrates heated to 700-750~$^\circ$C.  A Lambda Physik Compex 205 KrF laser operating at a wavelength of 248~nm has been used to supply 25~ns laser pulses at a constant pulse rate of 10~Hz and a fluence of between 4.0~Jcm$^{-2}$ and 7.5~Jcm$^{-2}$. Film structural quality was checked by RHEED and XRD, which showed that the films grew crystalline, aligned with the substrate and with an out-of-plane lattice parameter of 8.063~\AA{}, in close agreement with the bulk value of 8.062~\AA{}~\cite{borges_1999}.  Full details of film growth, RHEED and XRD are reported elsewhere~\cite{farrell_2017}.  

Magnetization has been measured using a Quantum Design MPMS XL SQUID magnetometer, at temperatures from 2-400~K, and in applied fields of up to 70~kOe ($\sim$5.6$\times$10$^6$~Am$^{-1}$) applied in the plane of the films.  Magnetization values have been calculated using the Ba$_2$FeMoO$_6$ formula unit volume of 1.285$\times$10$^{-22}$~cm$^3$.  Transport measurements were made with a Quantum Design PPMS at temperatures from 2.5-400~K and in fields up to 90~kOe ($\sim$7.2$\times$10$^6$~Am$^{-1}$).  The resistivity rotator rod option was used, enabling the films to be rotated for measurements with field applied both in-plane and out-of-plane.  Films were also mounted vertically on the rotator rod puck in order to rotate the field in plane for anisotropic magnetoresistance measurements. A thin layer of GE varnish was used to thermally anchor the samples to the PPMS pucks and electrical contacts were made to the films with Cu wires and EPO-TEK$^{\textregistered}$ silver epoxy.    

\section{Results and Analysis}
\subsection{Magnetization and resistivity}
\begin{figure}[htp]  \vspace{-0.5 cm}

\includegraphics[width=9cm]{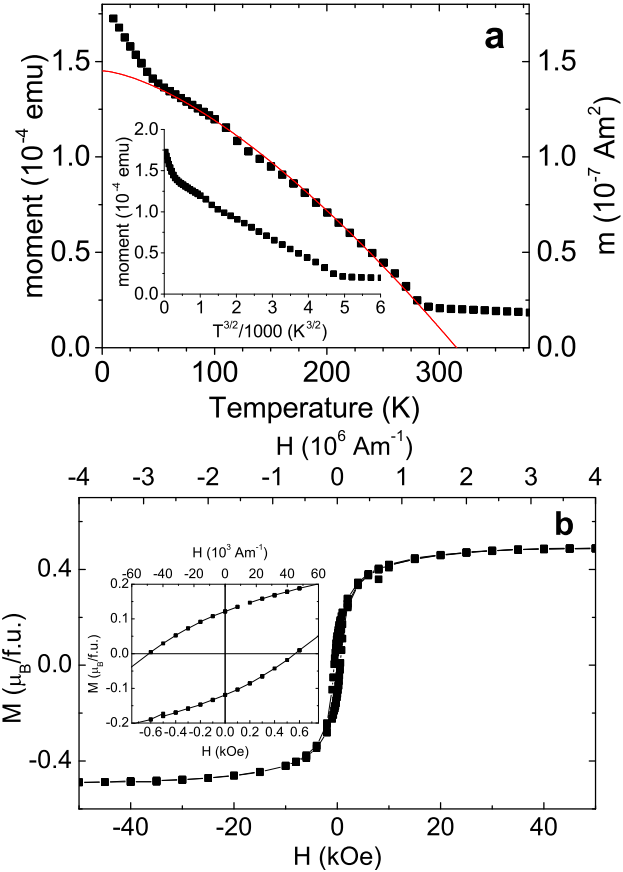}
\caption{(a) Remanent magnetization vs temperature of a Ba$_2$FeMoO$_6$ thin film, measured after field cooling in 100~Oe ($\sim8.0\times10^3$ Am$^{-1}$), with an extrapolated Curie temperature $\sim$305~K.  The line is a fit to a $T^{3/2}$ dependence. (b) Magnetization vs field at 10~K. 
\label{film_Mag}}
\vspace{-0.5 cm}
\end{figure}
\bigskip

The Ba$_2$FeMoO$_6$ films are ferromagnetic up to near room temperature, with temperature and field-dependent magnetization measurements shown in Figure~\ref{film_Mag}.  The field loop data have been corrected for the negative linear diamagnetic contribution from the substrate which dominates at high fields.  A low magnetic background is visible in the temperature-dependent magnetization above $\sim$300~K, possibly due to a small Fe impurity below the XRD detection limit.  This impurity is often seen in double perovskite oxides~\cite{tovar_2002,navarro_2003} due to the stability of some of the precursor phases in the preparation of bulk material.  The magnetization is proportional to $T^{3/2}$, fitting the well-known Bloch law of magnetization decay due to the excitation of magnons in an isotropic ferromagnet~\cite{blundell_2001}. An extrapolation of the $T^{3/2}$ dependence to zero moment results in an approximate Curie temperature of $\sim$305~K.  The saturation magnetization $M$\textsubscript{S} at 10~K of the film shown in Fig.~\ref{film_Mag} is 0.48~$\mu$\textsubscript{B}, with the largest $M$\textsubscript{S} of our films 0.75 $\mu$\textsubscript{B} (not shown), both considerably lower than the 3.72~$\mu$\textsubscript{B} in the optimized bulk target. However, the lower values of our films are close to those achieved in the only other published studies on PLD thin films~\cite{kim_magnetic_2012,kim_effects_2015}.  

\begin{figure}
\includegraphics[width=9cm]{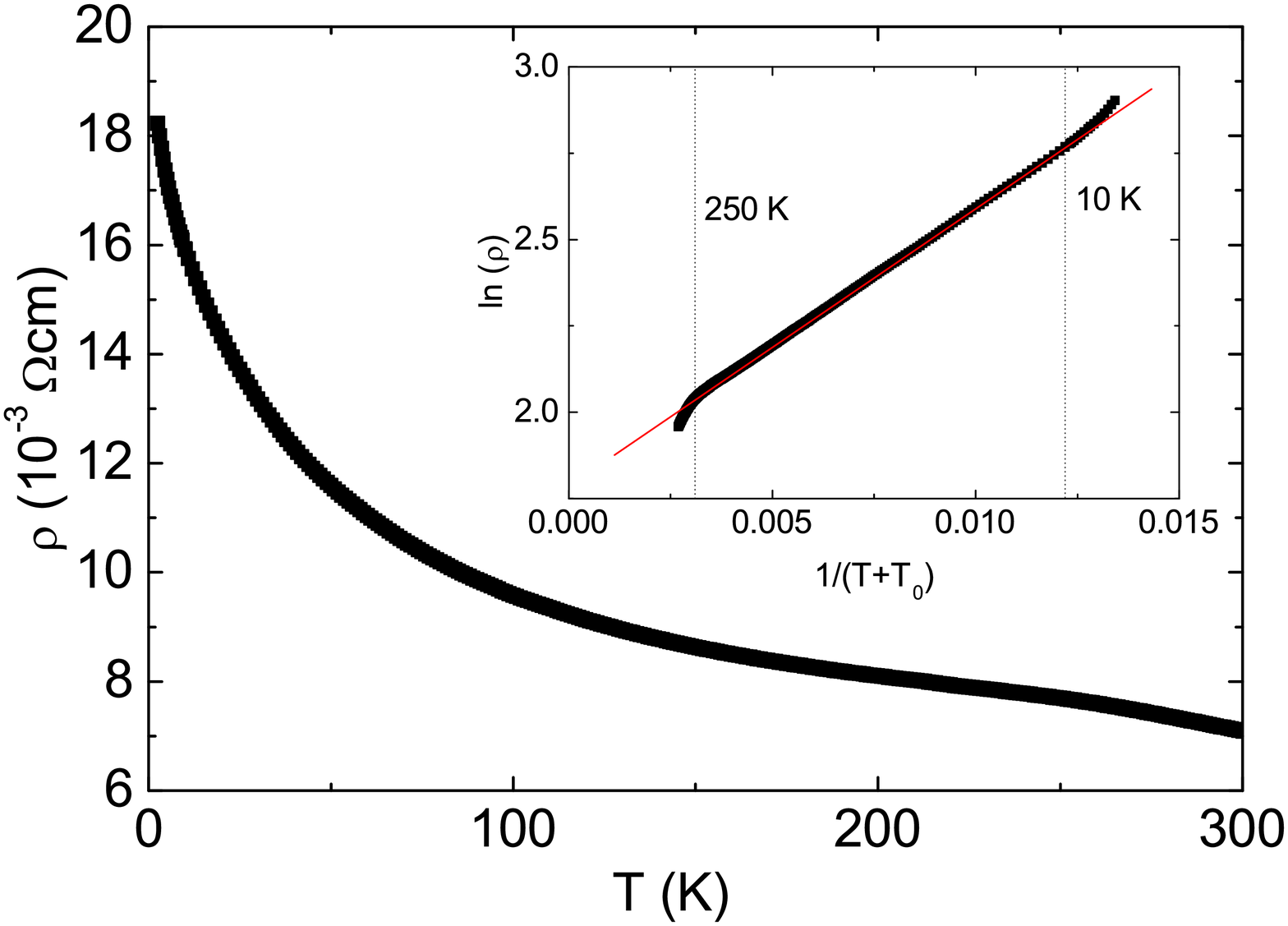}
\caption{Resistivity vs temperature of a Ba$_2$FeMoO$_6$ film in zero applied field. Inset: test of fluctuation-assisted tunnelling model of conduction.
\label{film_RT}}\vspace{-0.5 cm}
\end{figure}

\bigskip
The resistivity is shown in Figure~\ref{film_RT}, with a negative temperature coefficient implying a non-metallic conduction behavior.  However, the resistivity change from 2.5~K to 300~K is less than one order of magnitude, and the resistivity values of our films, at $\sim$10$^{-2}$~$\Omega$cm, agree with the values of 10$^{-4}$-10$^{-2}$~$\Omega$cm found elsewhere in bulk pellets of BFMO with metallic conduction~\cite{pandey_magnetic_2009,park_2001}. We have attempted to fit the data with $\rho$ $\propto$ $e^{(\frac{a}{T})^n}$, for 
$n=1,\frac{1}{2},\frac{1}{3},\frac{1}{4}$ as described by typical models for the temperature-dependent resistivity of semiconductors~\cite{mott_1979,shklovskii_1984}, but none fit over more than a small range of temperatures. On the other hand, the inset of Figure~\ref{film_RT} shows that the temperature-dependence conforms over nearly the full temperature range measured to the fluctuation-assisted tunnelling (FAT) conduction model, where resistivity is dominated by tunnelling of carriers between conducting grains. In this model the conducting grains are separated by energy barriers~\cite{sheng_1980} with a barrier height that varies due to thermal fluctuations, resulting in a resistivity given by  

\begin{equation}
\rho = \rho_0 e^\frac{T_1}{T+T_0}.
\label{rho_FAT_eqn}
\end{equation}

We estimate the physical thickness of the barrier from the ratio $\frac{T_1}{T_0}$= $\frac{\pi\chi w}{2}$, where 
$\chi=\sqrt{\frac{2mV_0}{\hbar^2}}$ is the reciprocal length of the wave function, $V_0$ is the barrier potential height and $w$ is the barrier thickness. The resistivity obeys the FAT model well up to approximately 250~K, which gives an estimate of where the thermal energy $k$\textsubscript{B}$T$ overcomes the energy barrier, implying $V_0\sim$20~meV.  Taking the effective mass to be 1.7-3.3 from the similar compound Sr$_2$FeMoO$_6$~\cite{moritomo_2000,saitoh_2005,popuri_2015}, we then find that $\chi$ is between $9.4\times$$10^8$~m$^{-1}$ and $13\times$$10^8$~m$^{-1}$ and, using the fitted values of $T_1$=80.0 $\pm$ 0.1~K and $T_0$=72.3 $\pm$ 0.1~K, the barrier width 
$w$=0.5-0.7~nm, which compares well with values of the barrier width of 0.4-3.0~nm found from polycrystalline Sr$_2$FeMoO$_6$~\cite{fisher_2003,fisher_2006,fisher_2014}.  The average crystallite size in our BFMO films is $\sim$37~nm, calculated from the $K_\alpha$ (004) BFMO reflection full-width at half-maximum of 0.006 rad (0.37$^\circ$) with the Scherrer formula and the measured film XRD spectrum~\cite{farrell_2017}.  The film is therefore comprised mostly of bulk-like conducting BFMO crystallites, with intergranular conduction occurring through the thin tunnelling barriers of possibly disordered or sub-stoichiometric material.  

\begin{figure}[htp]  \vspace{-0.0 cm}


\includegraphics[width=8.5cm]{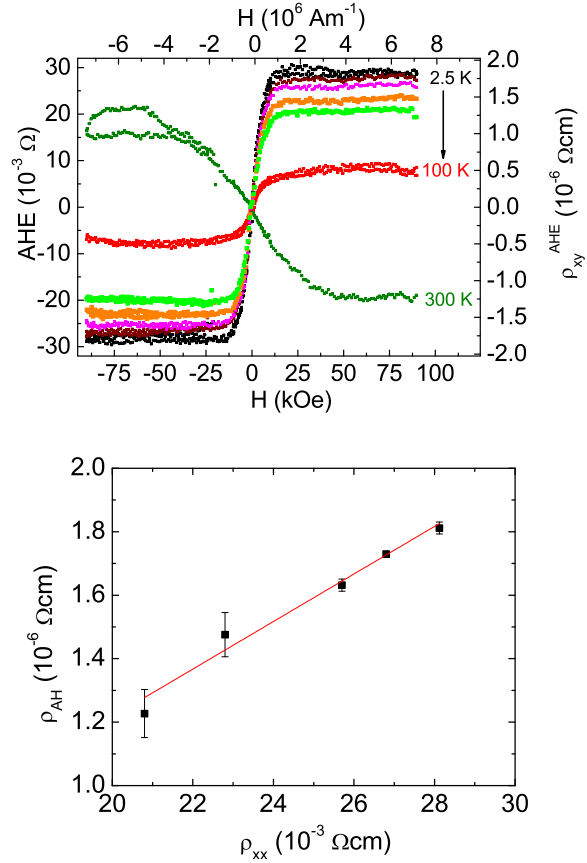}
\caption{(color online) Top: Anomalous Hall effect at various temperatures.  The high-field linear ordinary Hall effect component has been subtracted to better show the saturating anomalous component.  The sign changes from positive to negative between 100~K and 300~K. Bottom: Linear dependence of anomalous Hall effect on $\rho_{xx}$ below 50~K.}
\label{film_AHE}
\end{figure}

\subsection{Hall effect}

Figure~\ref{film_AHE} shows the anomalous Hall (AHE) effect of the BFMO film measured at various temperatures, with the linear high field component subtracted. The linear component, due to the ordinary Hall effect, is -7.03$\times$$10^{-11}$~$\Omega$cm/Oe at 2.5~K, corresponding to an electron carrier density of 8.89$\times$$10^{20}$~cm$^{-3}$ or 0.11~electrons per BFMO formula unit, and with a mobility of 0.39~cm$^2$/Vs. The carrier concentration agrees with values previously reported by Kim \textit{et al.} for BFMO films~\cite{kim_magnetic_2012} and is an order of magnitude lower than reported for SFMO thin films~\cite{westerburg_hall_2000,westerburg_epitaxy_2000}. At 100~K, the carrier concentration has increased to 0.56~electrons/f.u. and mobility has decreased to 0.15~cm$^2$/Vs. The anomalous Hall resistivity was calculated from the corrected Hall voltage using $\rho_{xy}=\frac{V_{\textrm{H}}t}{I}$, where $V_{\textrm{H}}$ is the Hall voltage, $t$ is the thickness of 630~nm and $I$ is the measurement current 10~mA. 
\bigskip

At low temperatures the sign of the anomalous Hall resistivity is positive, reaching 1.8~$\mu\Omega$cm at 2~K, dropping steadily as temperature increases up to 100~K. The anomalous Hall resistivity is linearly dependent on resistivity $\rho_{xx}$ as shown in the bottom part of Figure~\ref{film_AHE}, implying the dominance of the skew scattering mechanism of the anomalous Hall effect~\cite{nagaosa_2010}. An interesting phenomenon is the change of sign of the AHE from positive to negative between 100~K and 300~K.  While we cannot immediately rule out that the origin of this negative AHE is the weakly magnetic component seen in Figure~\ref{film_Mag}a, our measured AHE is of the opposite sign to that seen in both Fe thin films~\cite{tian_2009} and nanoparticles~\cite{aronzon_2000}. 
To understand the change of sign, we turn to Yanagihara \textit{et al.}, who discussed two contributions to the scattering that gives rise to the AHE in SFMO single crystals~\cite{yanagihara_magnetotransport_2001}. These two contributions are scattering from magnetic and from non-magnetic impurities, which have opposite signs in the scattering amplitude. As in SFMO single crystals, at low temperatures our BFMO films have a positive sign of AHE, implying that the AHE is dominated by the magnetic scattering component. Nearer $T$\textsubscript{C}, the change in sign of the AHE to negative shows that the magnetic scattering contribution weakens and the non-magnetic scattering dominates.  

\begin{figure}
\includegraphics[width=9.5cm]{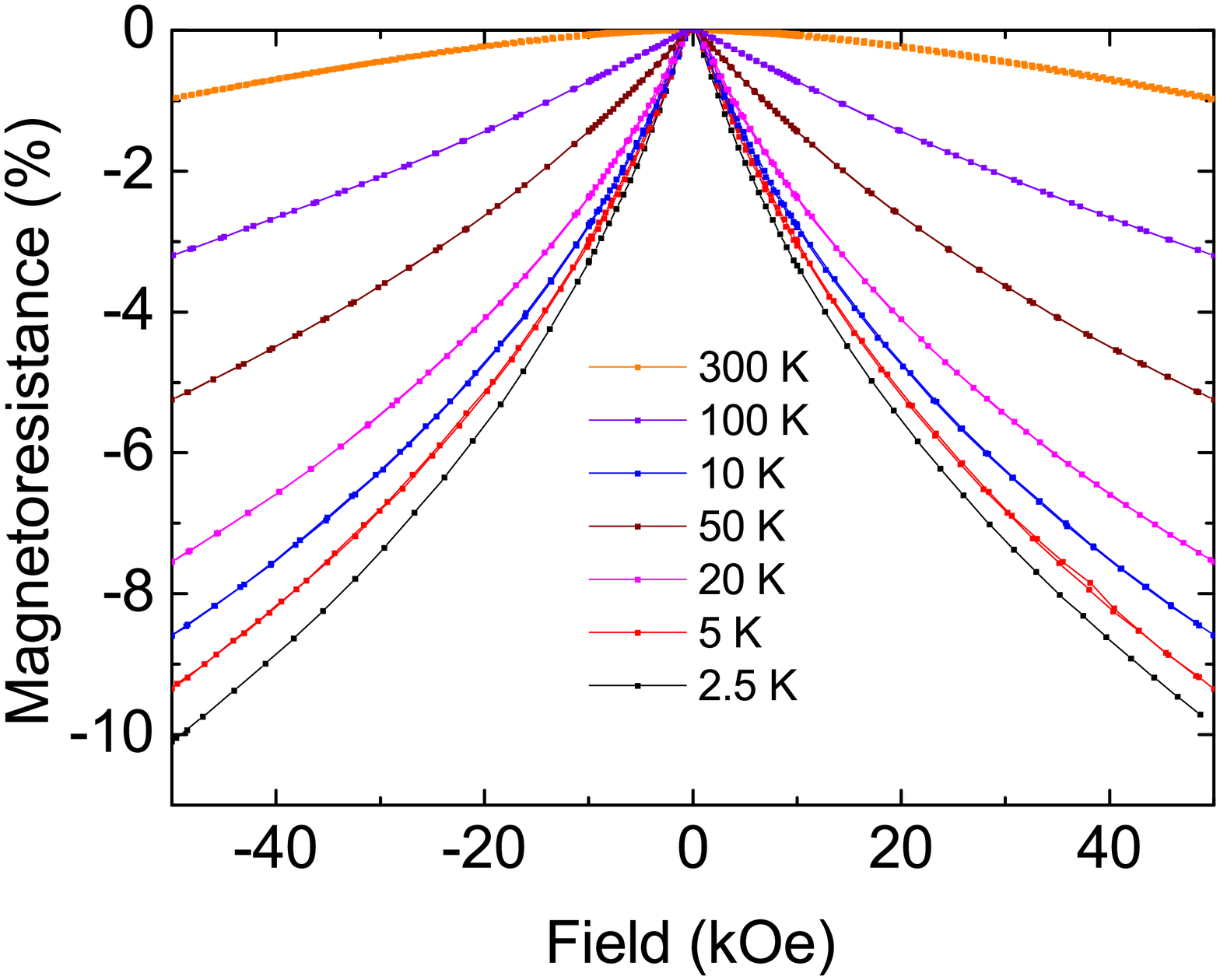}
\caption{(color online) Longitudinal magnetoresistance of Ba$_2$FeMoO$_6$ at various temperatures. Lines are fits to equation~\ref{MReqn}. 
\label{film_MR}}\vspace{-0.5 cm}
\end{figure}

\subsection{Magnetoresistance}
The longitudinal magnetoresistance (longitudinal MR), i.e., the field-dependent change of resistance measured parallel to the direction of the applied current, is shown in Figure~\ref{film_MR}. The longitudinal MR is negative, and below $T$\textsubscript{C} the MR becomes stronger as temperature decreases, reaching a maximum of -10\% at 50~kOe ($\sim$4.0$\times$$10^6$~Am$^{-1}$) and 2.5~K. A notable feature of the longitudinal MR is that it does not saturate, even at the lowest temperatures and highest fields. This is again different to the behavior of the magnetization of Fig.~\ref{film_Mag}, which has already reached saturation by approximately 20~kOe. The non-saturating MR is a familiar phenomenon to granular oxide films with strong spin polarization such as CrO$_2$ and CrO$_2$/Cr$_x$O$_y$~\cite{bajpai_2007} and Fe$_3$O$_4$-Pt~\cite{cheng_2011}, all of which also exhibit the FAT conduction mechanism. Similar magnetoresistance has been reported in bulk (Sr,Ba)$_2$FeMoO$_6$~\cite{serrate_2005,stephen_magnetic_2013} and has been successfully modelled as

\begin{equation}
\mathrm{MR}=-\frac{P^2m(H)^2}{1+P^2m(H)^2}~+~AH , 
\label{MReqn}
\end{equation}

\noindent where $P$ is the spin polarization, $H$ is the magnetic field, $m(H)$ is the field-dependent magentization and $A$ is a constant. The first term is due to tunnelling magnetoresistance, where $m(H)$ takes the form used for spin glasses with a weak anisotropy field, and the second term is linear in field due to spin disorder. However, in the bulk samples of refs ~\cite{serrate_2005} and ~\cite{stephen_magnetic_2013}, the temperature coefficient of resistivity was positive, unlike the semiconductor-like negative coefficient of resistivity in our thin films. While the data of Figure~\ref{film_MR} can be well fit to equation~\ref{MReqn}, the fitted spin polarization $P$ does not exceed 10\% at 2.5~K, much lower than expected for BFMO.

\section{Discussion}

From Figure~\ref{film_Mag}a, the saturation magnetization of the BFMO films is only $\sim$20\% of the theoretical maximum of 4~$\mu$\textsubscript{B}/f.u., which could most simply be explained by a large fraction of the film being magnetically 'dead', due to defects or composition variations. It is tempting to attribute the tunnel grain boundaries identified by the temperature-dependent resistivity measurements to such a magnetically dead component. However, with a grain diameter of 37~nm extracted from the XRD, the grain boundary material would need to form a shell 10~nm thick around each grain to account for the reduced magnetization, more than an order of magnitude larger than the width of the defected region indicated by the resistivity measurements. In double perovskites, typically the source of low saturation magnetization is anti-site disorder (ASD)~\cite{ogale_octahedral_1999,huang_2006}, with the reduced magnetization primarily arising from an antiferromagnetic coupling of the site-swapped Fe ions. In addition, ASD reduces $T$\textsubscript{C}~\cite{wang_2013}, with $\sim$20\% ASD reducing $T$\textsubscript{C} by $\sim$25~K in SFMO. We see a similarly lower $T$\textsubscript{C} in our BFMO films, 305~K compared to $T$\textsubscript{C}=315~K in the low-ASD BFMO pellet. However, such a large ASD should also show up as an increased resistivity compared to bulk, low ASD material. This was the case for Kim et al.~\cite{kim_magnetic_2012}, who reported a resistivity of 200-600~$\mu\Omega$cm in their low $M$\textsubscript{S} BFMO films. The resistivity of our films is an order of magnitude less and is therefore inconsistent with the presence of high ASD
.  However, our films clearly obey the fluctuation-assisted tunnelling conduction model of metallic grains separated by tunnel barriers, indicating that defected material is present as grain boundaries. 
 
\bigskip
To explain these results, we hypothesize that in our films the ASD is spatially correlated rather than more uniformly distributed. Calculations have shown that short-range ordering in double perovskites is robust, and the formation of clustered patches of ASD into anti-phase boundaries (APBs)~\cite{goodenough_comment_2000} is favoured over homogeneous ASD~\cite{meneghini_nature_2009}. APBs separate magnetic domains that are aligned anti-parallel, and when the magnetic domains are small enough, even large magnetic fields may be insufficient to align them in the same sense. The domain walls extend only a few atoms from the point defects that create the APBs and this size is a close match to the tunnel barrier width extracted from the FAT conductivity behaviour. Correlated ASD in the form of APBs is thus a likely explanation for the lower saturation magnetization and magnetotransport properties of our films, while explaining also the observed temperature-dependent resistivity. The point defected APB regions are too small to be easily detected with conventional XRD, which indicates a structural coherence over 37~nm, a volume which covers many distinct magnetic domains and APB regions.

\begin{figure}
\includegraphics[width=9.5cm]{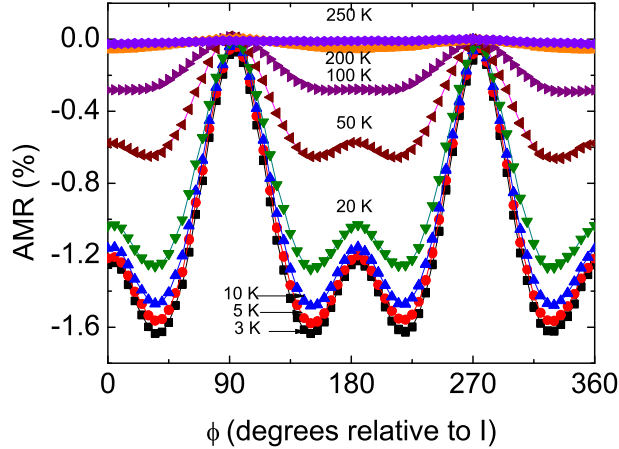}
\caption{(color online) Anisotropic magnetoresistance of Ba$_2$FeMoO$_6$ at 60~kOe ($\sim$4.8$\times$$10^6$~Am$^{-1}$) at various temperatures. $\phi$ indicates the angle of the in-plane magnetic field relative to the applied current $I$. The AMR is set to 0 at the angle where the resistance is a maximum ($\phi$=90$^{\circ}$)
\label{film_AMR_6T}}\vspace{-0.5 cm}
\end{figure}

\bigskip
To understand better the magnetotransport measurements, we measure the anisotropic magnetoresistance (AMR), shown in Figure~\ref{film_AMR_6T}. AMR is the change in resistance measured as the magnetic field is rotated in the plane of the film, in contrast to the longitudinal MR, which measures the resistance change with a fixed field direction. The conventional theory of AMR in transition metal-based compounds involves scattering from itinerant electrons (usually $s$ states) to localized states with a magnetic moment (usually 3$d$ states)~\cite{mcguire_anisotropic_1975}. While BFMO has little to no $s$ electron occupation in the conduction band at the Fermi level~\cite{kanchana_electronic_2007}, the conduction band is mostly made up of states from itinerant Mo 4$d$ electrons~\cite{kobayashi_1998,szotek_electronic_2003,mohamed_musa_saad_first-principles_2012} hybridized with localized Fe 3$d$ orbitals~\cite{meetei_theory_2013}, such that the concept of AMR developed for itinerant electrons can also be applied to the AMR we observe in BFMO. 

\bigskip
Most notably, the AMR of Figure~\ref{film_AMR_6T} is negative for temperatures below $\sim$250~K, coinciding with the temperature where the AHE changes sign. A negative AMR is predicted to be an indicator of half-metallic conduction~\cite{kokado_anisotropic_2012,kokado_anisotropic_2013}, opposite to the ordinary positive AMR in transition metal ferromagnets and alloys~\cite{mcguire_anisotropic_1975}. This has been supported by measurements of negative AMR in half-metallic Heusler thin films~\cite{yang_anisotropic_2012,du_half-metallicity_2013,sakuraba_quantitative_2014,bosu_enhancement_2016}. Hence, the negative AMR of Figure~\ref{film_AMR_6T} indicates the BFMO is half-metallic below $\sim$250-300~K.

\bigskip 
    
The 
low value of spin polarization derived from the longitudinal MR measurements appears incompatible with the indication of half-metallicity from AMR measurements. However, the origin of the AMR is different from that of the longitudinal MR. At in-plane fields larger than $\sim$30~kOe, the magnetic grains in the films are saturated and the lack of change in the magnetization up to 50~kOe in Figure~\ref{film_Mag}b shows there can be no effect of rotating the magnetic field on the magnetoresistance due to inter-grain transport, as the magnetic moment of the grains should rotate to follow the field direction. 
The AMR at large fields is instead due to intra-grain scattering, and probes the orbital electronic states. Hence, the AMR is sensitive to the electronic structure \textit{within} the magnetic grains, whereas the longitudinal MR depends on tunnelling \textit{between} grains. The negative sign of the AMR indicates that the magnetic grains are half-metallic, as predicted for BFMO. As implied by the resistivity measurements of Figure~\ref{film_RT}, in between the grains are correlated APBs made of defected BFMO material. The presence of APBs is supported by the appearance in Figure~\ref{film_AMR_6T} of a four-fold component in the AMR, below $\sim$100~K. A similar four-fold component associated with APBs has been noticed in epitaxial Fe$_3$O$_4$ thin films~\cite{li_fourfold_2010,li_origin_2010,jin_magnetocrystalline_2012}. Spin polarized electrons from the half-metallic BFMO grains experience extra scattering when tunnelling out through the domain wall APB regions, leading to the tunnel-barrier resistivity and low spin polarization. The non-saturation of the longitudinal MR confirms that the APBs are unable to reach full alignment, even at 5~K and 60~kOe.

\bigskip
In summary, the temperature-dependent resistivity, low saturation magnetization and non-saturation of magnetoresistances show that defects, probably APBs, inhibit full bulk-like magnetic alignment of the films, which are broken into small magnetic domains, while the double perovskite structure as probed by conventional XRD is coherent over a longer scale. The APBs provide barriers to the spin polarized charge carriers that tunnel between metallic BFMO grains, which reduces the effective intergrain spin polarization. However, the volume of these barrier regions is small compared to the volume of the well-ordered BFMO grains, and the AMR and anomalous Hall effect measurements indicate half-metallic conduction in the BFMO thin films, which disappears around 250-300~K, close to the film $T$\textsubscript{C} of $\sim$305~K. This is an indication that a strong spin polarization is present in the grains, even in BFMO films that do not show the ideal bulk magnetic character. The main challenge now then is to optimize the grain boundaries in order to allow for the spin polarized carriers to tunnel out of half-metallic BFMO grains without significant spin scattering.  

\section{Conclusion}

Thin films of BFMO have been produced by PLD on lattice-matched substrates from an optimized bulk target with low anti-site disorder. As is often found for double perovskites in thin film form, the films have a magnetization lower than the bulk and a non-saturating magnetoresistance due to magnetically defected grain boundary regions, likely to be antiphase boundaries between otherwise conducting BFMO grains. However, transport measurements show signs of strongly spin polarized carriers to $\sim$250-300~K, where a change in sign of the anomalous Hall effect and anisotropic magnetoresistance occurs near $T$\textsubscript{C} when the magnetic ordering is lost. The spin polarized carriers originate from crystalline BFMO grains, with intergranular conduction occurring through the anti-phase boundaries that act as weak tunnel barriers to conduction. BFMO may thus be a good source of a strongly spin polarized current for use in thin film spintronic devices, even where defects at grain boundaries cause spin scattering and reduce the magnetization lower than the theoretical maximum value of 4~$\mu$\textsubscript{B}/f.u.  This is a promising sign for the prospective use of BFMO as a source of highly spin polarized carriers in spintronic devices, if the spin-scattering grain boundary material can be reduced.
\bigskip

We gratefully acknowledge funding and equipment support from the New Zealand Ministry of Business, Innovation and Employment (contract C08X0705), and the MacDiarmid Institute for Advanced Materials and Nanotechnology.
\bibliography{Granville_PRB_BFMO_films2_1_column_v2}

\begin{thebibliography}{70}%
	\makeatletter
	\providecommand \@ifxundefined [1]{%
		\@ifx{#1\undefined}
	}%
	\providecommand \@ifnum [1]{%
		\ifnum #1\expandafter \@firstoftwo
		\else \expandafter \@secondoftwo
		\fi
	}%
	\providecommand \@ifx [1]{%
		\ifx #1\expandafter \@firstoftwo
		\else \expandafter \@secondoftwo
		\fi
	}%
	\providecommand \natexlab [1]{#1}%
	\providecommand \enquote  [1]{``#1''}%
	\providecommand \bibnamefont  [1]{#1}%
	\providecommand \bibfnamefont [1]{#1}%
	\providecommand \citenamefont [1]{#1}%
	\providecommand \href@noop [0]{\@secondoftwo}%
	\providecommand \href [0]{\begingroup \@sanitize@url \@href}%
	\providecommand \@href[1]{\@@startlink{#1}\@@href}%
	\providecommand \@@href[1]{\endgroup#1\@@endlink}%
	\providecommand \@sanitize@url [0]{\catcode `\\12\catcode `\$12\catcode
		`\&12\catcode `\#12\catcode `\^12\catcode `\_12\catcode `\%12\relax}%
	\providecommand \@@startlink[1]{}%
	\providecommand \@@endlink[0]{}%
	\providecommand \url  [0]{\begingroup\@sanitize@url \@url }%
	\providecommand \@url [1]{\endgroup\@href {#1}{\urlprefix }}%
	\providecommand \urlprefix  [0]{URL }%
	\providecommand \Eprint [0]{\href }%
	\providecommand \doibase [0]{http://dx.doi.org/}%
	\providecommand \selectlanguage [0]{\@gobble}%
	\providecommand \bibinfo  [0]{\@secondoftwo}%
	\providecommand \bibfield  [0]{\@secondoftwo}%
	\providecommand \translation [1]{[#1]}%
	\providecommand \BibitemOpen [0]{}%
	\providecommand \bibitemStop [0]{}%
	\providecommand \bibitemNoStop [0]{.\EOS\space}%
	\providecommand \EOS [0]{\spacefactor3000\relax}%
	\providecommand \BibitemShut  [1]{\csname bibitem#1\endcsname}%
	\let\auto@bib@innerbib\@empty
	\bibitem [{\citenamefont {Kobayashi}\ \emph {et~al.}(1998)\citenamefont
		{Kobayashi}, \citenamefont {Kimura}, \citenamefont {Sawada}, \citenamefont
		{Terakura},\ and\ \citenamefont {Tokura}}]{kobayashi_1998}%
	\BibitemOpen
	\bibfield  {author} {\bibinfo {author} {\bibfnamefont {K.-I.}\ \bibnamefont
			{Kobayashi}}, \bibinfo {author} {\bibfnamefont {T.}~\bibnamefont {Kimura}},
		\bibinfo {author} {\bibfnamefont {H.}~\bibnamefont {Sawada}}, \bibinfo
		{author} {\bibfnamefont {K.}~\bibnamefont {Terakura}}, \ and\ \bibinfo
		{author} {\bibfnamefont {Y.~H.}\ \bibnamefont {Tokura}},\ }\href
	{http://www.nature.com/nature/journal/v395/n6703/abs/395675a0.html}
	{\bibfield  {journal} {\bibinfo  {journal} {Nature}\ }\textbf {\bibinfo
			{volume} {395}},\ \bibinfo {pages} {675} (\bibinfo {year}
		{1998})}\BibitemShut {NoStop}%
	\bibitem [{\citenamefont {Bibes}\ and\ \citenamefont
		{Barthelemy}(2007)}]{bibes_2007}%
	\BibitemOpen
	\bibfield  {author} {\bibinfo {author} {\bibfnamefont {M.}~\bibnamefont
			{Bibes}}\ and\ \bibinfo {author} {\bibfnamefont {A.}~\bibnamefont
			{Barthelemy}},\ }\href {\doibase 10.1109/TED.2007.894366} {\bibfield
		{journal} {\bibinfo  {journal} {IEEE Trans. Electron Devices}\ }\textbf
		{\bibinfo {volume} {54}},\ \bibinfo {pages} {1003} (\bibinfo {year}
		{2007})}\BibitemShut {NoStop}%
	\bibitem [{\citenamefont {Hirohata}\ \emph {et~al.}(2015)\citenamefont
		{Hirohata}, \citenamefont {Sukegawa}, \citenamefont {Yanagihara},
		\citenamefont {Zuti\c{c}}, \citenamefont {Seki}, \citenamefont {Mizukami},\
		and\ \citenamefont {Swaminathan}}]{hirohata_2015}%
	\BibitemOpen
	\bibfield  {author} {\bibinfo {author} {\bibfnamefont {A.}~\bibnamefont
			{Hirohata}}, \bibinfo {author} {\bibfnamefont {H.}~\bibnamefont {Sukegawa}},
		\bibinfo {author} {\bibfnamefont {H.}~\bibnamefont {Yanagihara}}, \bibinfo
		{author} {\bibfnamefont {I.}~\bibnamefont {Zuti\c{c}}}, \bibinfo {author}
		{\bibfnamefont {T.}~\bibnamefont {Seki}}, \bibinfo {author} {\bibfnamefont
			{S.}~\bibnamefont {Mizukami}}, \ and\ \bibinfo {author} {\bibfnamefont
			{R.}~\bibnamefont {Swaminathan}},\ }\href {\doibase
		10.1109/TMAG.2015.2457393} {\bibfield  {journal} {\bibinfo  {journal} {IEEE
				Trans. Magn.}\ }\textbf {\bibinfo {volume} {51}},\ \bibinfo {pages} {1}
		(\bibinfo {year} {2015})}\BibitemShut {NoStop}%
	\bibitem [{\citenamefont {Borges}\ \emph {et~al.}(1999)\citenamefont {Borges},
		\citenamefont {Thomas}, \citenamefont {Cullinan}, \citenamefont {Coey},
		\citenamefont {Suryanarayanan}, \citenamefont {Ben-Dor}, \citenamefont
		{Pinsard-Gaudart},\ and\ \citenamefont {Revcolevschi}}]{borges_1999}%
	\BibitemOpen
	\bibfield  {author} {\bibinfo {author} {\bibfnamefont {R.~P.}\ \bibnamefont
			{Borges}}, \bibinfo {author} {\bibfnamefont {R.~M.}\ \bibnamefont {Thomas}},
		\bibinfo {author} {\bibfnamefont {C.}~\bibnamefont {Cullinan}}, \bibinfo
		{author} {\bibfnamefont {J.~M.~D.}\ \bibnamefont {Coey}}, \bibinfo {author}
		{\bibfnamefont {R.}~\bibnamefont {Suryanarayanan}}, \bibinfo {author}
		{\bibfnamefont {L.}~\bibnamefont {Ben-Dor}}, \bibinfo {author} {\bibfnamefont
			{L.}~\bibnamefont {Pinsard-Gaudart}}, \ and\ \bibinfo {author} {\bibfnamefont
			{A.}~\bibnamefont {Revcolevschi}},\ }\href
	{http://iopscience.iop.org/0953-8984/11/40/104} {\bibfield  {journal}
		{\bibinfo  {journal} {J. Phys. Condens. Matter}\ }\textbf {\bibinfo {volume}
			{11}},\ \bibinfo {pages} {L445} (\bibinfo {year} {1999})}\BibitemShut
	{NoStop}%
	\bibitem [{\citenamefont {Szotek}\ \emph {et~al.}(2003)\citenamefont {Szotek},
		\citenamefont {Temmerman}, \citenamefont {Svane}, \citenamefont {Petit},\
		and\ \citenamefont {Winter}}]{szotek_electronic_2003}%
	\BibitemOpen
	\bibfield  {author} {\bibinfo {author} {\bibfnamefont {Z.}~\bibnamefont
			{Szotek}}, \bibinfo {author} {\bibfnamefont {W.~M.}\ \bibnamefont
			{Temmerman}}, \bibinfo {author} {\bibfnamefont {A.}~\bibnamefont {Svane}},
		\bibinfo {author} {\bibfnamefont {L.}~\bibnamefont {Petit}}, \ and\ \bibinfo
		{author} {\bibfnamefont {H.}~\bibnamefont {Winter}},\ }\href {\doibase
		10.1103/PhysRevB.68.104411} {\bibfield  {journal} {\bibinfo  {journal} {Phys.
				Rev. B}\ }\textbf {\bibinfo {volume} {68}},\ \bibinfo {pages} {104411}
		(\bibinfo {year} {2003})}\BibitemShut {NoStop}%
	\bibitem [{\citenamefont {Itoh}\ \emph {et~al.}(1996)\citenamefont {Itoh},
		\citenamefont {Ohta},\ and\ \citenamefont {Inaguma}}]{itoh_1996}%
	\BibitemOpen
	\bibfield  {author} {\bibinfo {author} {\bibfnamefont {M.}~\bibnamefont
			{Itoh}}, \bibinfo {author} {\bibfnamefont {I.}~\bibnamefont {Ohta}}, \ and\
		\bibinfo {author} {\bibfnamefont {Y.}~\bibnamefont {Inaguma}},\ }\href@noop
	{} {\bibfield  {journal} {\bibinfo  {journal} {Mater. Sci. Eng. B}\ }\textbf
		{\bibinfo {volume} {41}},\ \bibinfo {pages} {55} (\bibinfo {year}
		{1996})}\BibitemShut {NoStop}%
	\bibitem [{\citenamefont {Serrate}\ \emph {et~al.}(2007)\citenamefont
		{Serrate}, \citenamefont {Teresa},\ and\ \citenamefont
		{Ibarra}}]{serrate_double_2007}%
	\BibitemOpen
	\bibfield  {author} {\bibinfo {author} {\bibfnamefont {D.}~\bibnamefont
			{Serrate}}, \bibinfo {author} {\bibfnamefont {J.~M.~D.}\ \bibnamefont
			{Teresa}}, \ and\ \bibinfo {author} {\bibfnamefont {M.~R.}\ \bibnamefont
			{Ibarra}},\ }\href {\doibase 10.1088/0953-8984/19/2/023201} {\bibfield
		{journal} {\bibinfo  {journal} {J. Phys. Condens. Matter}\ }\textbf {\bibinfo
			{volume} {19}},\ \bibinfo {pages} {023201} (\bibinfo {year}
		{2007})}\BibitemShut {NoStop}%
\bibitem [{\citenamefont {Retuerto}\ \emph {et~al.}(2004)\citenamefont {Retuerto},
	\citenamefont {Alonso}, \citenamefont {Mart\'{i}nez-Lope}, \citenamefont {Mart\'{i}nez},\ and\ \citenamefont
	{\citenamefont {Garc\'{i}a-Hern\'{a}ndez}}}]{retuerto_2004}%
\BibitemOpen
\bibfield  {author} {\bibinfo {author} {\bibfnamefont {M.}~\bibnamefont
		{Retuerto}}, \bibinfo {author} {\bibfnamefont {J.~A.}~\bibnamefont {Alonso}}, \bibinfo
	{author} {\bibfnamefont {M.~J.}\ \bibnamefont {Mart\'{i}nez-Lope}}, \bibinfo
	{author} {\bibfnamefont {J.~L.}\ \bibnamefont {Mart\'{i}nez}}\ and\ \bibinfo {author}
	{\bibfnamefont {M.}\ \bibnamefont {Garc\'{i}a-Hern\'{a}ndez}},\ }\href@noop {} {\bibfield
	{journal} {\bibinfo  {journal} {Appl. Phys. Lett.}\ }\textbf {\bibinfo
		{volume} {85}},\ \bibinfo {pages} {266} (\bibinfo {year}
	{2004})}\BibitemShut {NoStop}%
\bibitem [{\citenamefont {Huang}\ \emph {et~al.}(2005)\citenamefont {Huang},
	\citenamefont {Lind\'{e}n}, \citenamefont {Yamauchi},\ and\ \citenamefont
	{\citenamefont {Karppinen}}}]{huang_2005}%
\BibitemOpen
\bibfield  {author} {\bibinfo {author} {\bibfnamefont {Y.~H.}~\bibnamefont
		{Huang}}, \bibinfo {author} {\bibfnamefont {J.}~\bibnamefont {Lind\'{e}n}}, \bibinfo
	{author} {\bibfnamefont {H.}\ \bibnamefont {Yamauchi}}\ and\ \bibinfo {author}
	{\bibfnamefont {M.}\ \bibnamefont {Karppinen}},\ }\href@noop {} {\bibfield
	{journal} {\bibinfo  {journal} {Appl. Phys. Lett.}\ }\textbf {\bibinfo
		{volume} {86}},\ \bibinfo {pages} {072510} (\bibinfo {year}
	{2005})}\BibitemShut {NoStop}%
\bibitem [{\citenamefont {Aldica}\ \emph {et~al.}(2007)\citenamefont {Aldica},
		\citenamefont {Plapcianu}, \citenamefont {Badica}, \citenamefont {Valsangiacom},\ and\ \citenamefont
		{Stoica}}]{aldica_2007}%
	\BibitemOpen
	\bibfield  {author} {\bibinfo {author} {\bibfnamefont {G.}~\bibnamefont
			{Aldica}}, \bibinfo {author} {\bibfnamefont {C.}~\bibnamefont {Plapcianu}}, \bibinfo
		{author} {\bibfnamefont {P.}\ \bibnamefont {Badica}}, \bibinfo {author} {\bibfnamefont {L.}~\bibnamefont {Valsangiacom}},\ and\ \bibinfo {author}
		{\bibfnamefont {L.}\ \bibnamefont {Stoica}},\ }\href@noop {} {\bibfield
		{journal} {\bibinfo  {journal} {J. Magn. Magn. Mater.}\ }\textbf {\bibinfo
			{volume} {311}},\ \bibinfo {pages} {665} (\bibinfo {year}
		{2007})}\BibitemShut {NoStop}%
\bibitem [{\citenamefont {S\'{a}nchez}\ \emph {et~al.}(2004)\citenamefont
		{S\'{a}nchez}, \citenamefont {Garc\'{i}a-Hern\'{a}ndez}, \citenamefont
		{Auth},\ and\ \citenamefont {Jakob}}]{sanchez_structural_2004}%
	\BibitemOpen
	\bibfield  {author} {\bibinfo {author} {\bibfnamefont {D.}~\bibnamefont
			{S\'{a}nchez}}, \bibinfo {author} {\bibfnamefont {M.}~\bibnamefont
			{Garc\'{i}a-Hern\'{a}ndez}}, \bibinfo {author} {\bibfnamefont
			{N.}~\bibnamefont {Auth}}, \ and\ \bibinfo {author} {\bibfnamefont
			{G.}~\bibnamefont {Jakob}},\ }\href {\doibase 10.1063/1.1774244} {\bibfield
		{journal} {\bibinfo  {journal} {J. Appl. Phys.}\ }\textbf {\bibinfo {volume}
			{96}},\ \bibinfo {pages} {2736} (\bibinfo {year} {2004})}\BibitemShut
	{NoStop}%
	\bibitem [{\citenamefont {Di~Trolio}\ \emph {et~al.}(2006)\citenamefont
		{Di~Trolio}, \citenamefont {Larciprete}, \citenamefont {Marotta},
		\citenamefont {Testa},\ and\ \citenamefont
		{Fiorani}}]{di_trolio_microstructure_2006}%
	\BibitemOpen
	\bibfield  {author} {\bibinfo {author} {\bibfnamefont {A.}~\bibnamefont
			{Di~Trolio}}, \bibinfo {author} {\bibfnamefont {R.}~\bibnamefont
			{Larciprete}}, \bibinfo {author} {\bibfnamefont {V.}~\bibnamefont {Marotta}},
		\bibinfo {author} {\bibfnamefont {A.~M.}\ \bibnamefont {Testa}}, \ and\
		\bibinfo {author} {\bibfnamefont {D.}~\bibnamefont {Fiorani}},\ }\href
	{\doibase 10.1002/pssc.200567113} {\bibfield  {journal} {\bibinfo  {journal}
			{Phys. Status Solidi C}\ }\textbf {\bibinfo {volume} {3}},\ \bibinfo {pages}
		{3229} (\bibinfo {year} {2006})}\BibitemShut {NoStop}%
	\bibitem [{\citenamefont {Wang}\ \emph {et~al.}(2006)\citenamefont {Wang},
		\citenamefont {Pan}, \citenamefont {Zhang}, \citenamefont {Lian},\ and\
		\citenamefont {Xiong}}]{wang_double-perovskite_2006}%
	\BibitemOpen
	\bibfield  {author} {\bibinfo {author} {\bibfnamefont {S.}~\bibnamefont
			{Wang}}, \bibinfo {author} {\bibfnamefont {H.}~\bibnamefont {Pan}}, \bibinfo
		{author} {\bibfnamefont {X.}~\bibnamefont {Zhang}}, \bibinfo {author}
		{\bibfnamefont {G.}~\bibnamefont {Lian}}, \ and\ \bibinfo {author}
		{\bibfnamefont {G.}~\bibnamefont {Xiong}},\ }\href {\doibase
		10.1063/1.2189009} {\bibfield  {journal} {\bibinfo  {journal} {Appl. Phys.
				Lett.}\ }\textbf {\bibinfo {volume} {88}},\ \bibinfo {pages} {121912}
		(\bibinfo {year} {2006})}\BibitemShut {NoStop}%
	\bibitem [{\citenamefont {Song}\ \emph {et~al.}(2008)\citenamefont {Song},
		\citenamefont {Park}, \citenamefont {Park},\ and\ \citenamefont
		{Jeong}}]{song_double-perovskite_2008}%
	\BibitemOpen
	\bibfield  {author} {\bibinfo {author} {\bibfnamefont {J.}~\bibnamefont
			{Song}}, \bibinfo {author} {\bibfnamefont {B.-G.}\ \bibnamefont {Park}},
		\bibinfo {author} {\bibfnamefont {J.-H.}\ \bibnamefont {Park}}, \ and\
		\bibinfo {author} {\bibfnamefont {Y.~H.}\ \bibnamefont {Jeong}},\ }\href@noop
	{} {\bibfield  {journal} {\bibinfo  {journal} {J. Korean Phys. Soc.}\
		}\textbf {\bibinfo {volume} {53}},\ \bibinfo {pages} {1084} (\bibinfo {year}
		{2008})}\BibitemShut {NoStop}%
	\bibitem [{\citenamefont {Hauser}\ \emph {et~al.}(2011)\citenamefont {Hauser},
		\citenamefont {Williams}, \citenamefont {Ricciardo}, \citenamefont {Genc},
		\citenamefont {Dixit}, \citenamefont {Lucy}, \citenamefont {Woodward},
		\citenamefont {Fraser},\ and\ \citenamefont {Yang}}]{hauser_unlocking_2011}%
	\BibitemOpen
	\bibfield  {author} {\bibinfo {author} {\bibfnamefont {A.~J.}\ \bibnamefont
			{Hauser}}, \bibinfo {author} {\bibfnamefont {R.~E.~A.}\ \bibnamefont
			{Williams}}, \bibinfo {author} {\bibfnamefont {R.~A.}\ \bibnamefont
			{Ricciardo}}, \bibinfo {author} {\bibfnamefont {A.}~\bibnamefont {Genc}},
		\bibinfo {author} {\bibfnamefont {M.}~\bibnamefont {Dixit}}, \bibinfo
		{author} {\bibfnamefont {J.~M.}\ \bibnamefont {Lucy}}, \bibinfo {author}
		{\bibfnamefont {P.~M.}\ \bibnamefont {Woodward}}, \bibinfo {author}
		{\bibfnamefont {H.~L.}\ \bibnamefont {Fraser}}, \ and\ \bibinfo {author}
		{\bibfnamefont {F.}~\bibnamefont {Yang}},\ }\href {\doibase
		10.1103/PhysRevB.83.014407} {\bibfield  {journal} {\bibinfo  {journal} {Phys.
				Rev. B}\ }\textbf {\bibinfo {volume} {83}},\ \bibinfo {pages} {014407}
		(\bibinfo {year} {2011})}\BibitemShut {NoStop}%
	\bibitem [{\citenamefont {Shinde}\ \emph
		{et~al.}(2003{\natexlab{a}})\citenamefont {Shinde}, \citenamefont {Ogale},
		\citenamefont {Greene}, \citenamefont {Venkatesan}, \citenamefont {Tsoi},
		\citenamefont {Cheong},\ and\ \citenamefont
		{Millis}}]{shinde_2003}%
	\BibitemOpen
	\bibfield  {author} {\bibinfo {author} {\bibfnamefont {S.~R.}~\bibnamefont
			{Shinde}}, \bibinfo {author} {\bibfnamefont {S.~B.}~\bibnamefont {Ogale}},
		\bibinfo {author} {\bibfnamefont {R.~L.}~\bibnamefont {Greene}}, \bibinfo
		{author} {\bibfnamefont {T.}~\bibnamefont {Venkatesan}}, \bibinfo {author}
		{\bibfnamefont {K.}~\bibnamefont {Tsoi}}, \bibinfo {author} {\bibfnamefont
			{S.-W.}~\bibnamefont {Cheong}}, \ and\ \bibinfo {author} {\bibfnamefont
			{A.~J.}~\bibnamefont {Millis}},\ }\href {\doibase 10.1063/1.1533831}
	{\bibfield  {journal} {\bibinfo  {journal} {J. Appl. Phys.}\ }\textbf {\bibinfo
			{volume} {93}},\ \bibinfo {pages} {1605} (\bibinfo {year}
		{2003}{\natexlab{a}})}\BibitemShut {NoStop}%
	\bibitem [{\citenamefont {Deniz}\ \emph
		{et~al.}(2017{\natexlab{a}})\citenamefont {Deniz}, \citenamefont {Preziosi},
		\citenamefont {Alexe},\ and\ \citenamefont
		{Hesse}}]{deniz_2017}%
	\BibitemOpen
	\bibfield  {author} {\bibinfo {author} {\bibfnamefont {H.}~\bibnamefont
			{Deniz}}, \bibinfo {author} {\bibfnamefont {D.}~\bibnamefont {Preziosi}},
		\bibinfo {author} {\bibfnamefont {M.}~\bibnamefont {Alexe}}, \ and\ \bibinfo {author} {\bibfnamefont
			{D.}~\bibnamefont {Hesse}},\ }\href {\doibase 10.1063/1.4973878}
	{\bibfield  {journal} {\bibinfo  {journal} {J. Appl. Phys.}\ }\textbf {\bibinfo
			{volume} {121}},\ \bibinfo {pages} {023906} (\bibinfo {year}
		{2017}{\natexlab{a}})}\BibitemShut {NoStop}%
	\bibitem [{\citenamefont {Kumar}\ \emph
		{et~al.}(2014{\natexlab{a}})\citenamefont {Kumar}, \citenamefont {Misra},
		\citenamefont {Kotnala}, \citenamefont {Gaur}, \citenamefont {Rawat},
		\citenamefont {Choudhary},\ and\ \citenamefont
		{Katiyar}}]{kumar_polycrystalline_2014}%
	\BibitemOpen
	\bibfield  {author} {\bibinfo {author} {\bibfnamefont {N.}~\bibnamefont
			{Kumar}}, \bibinfo {author} {\bibfnamefont {P.}~\bibnamefont {Misra}},
		\bibinfo {author} {\bibfnamefont {R.}~\bibnamefont {Kotnala}}, \bibinfo
		{author} {\bibfnamefont {A.}~\bibnamefont {Gaur}}, \bibinfo {author}
		{\bibfnamefont {R.}~\bibnamefont {Rawat}}, \bibinfo {author} {\bibfnamefont
			{R.}~\bibnamefont {Choudhary}}, \ and\ \bibinfo {author} {\bibfnamefont
			{R.}~\bibnamefont {Katiyar}},\ }\href {\doibase 10.1016/j.matlet.2013.11.113}
	{\bibfield  {journal} {\bibinfo  {journal} {Mater. Lett.}\ }\textbf {\bibinfo
			{volume} {118}},\ \bibinfo {pages} {200} (\bibinfo {year}
		{2014}{\natexlab{a}})}\BibitemShut {NoStop}%
	\bibitem [{\citenamefont {Bibes}\ \emph {et~al.}(2003)\citenamefont {Bibes},
		\citenamefont {Bouzehouane}, \citenamefont {Barth\'{e}l\'{e}my},
		\citenamefont {Besse}, \citenamefont {Fusil}, \citenamefont {Bowen},
		\citenamefont {Seneor}, \citenamefont {Carrey}, \citenamefont {Cros},
		\citenamefont {Vaurès}, \citenamefont {Contour},\ and\ \citenamefont
		{Fert}}]{bibes_tunnel_2003}%
	\BibitemOpen
	\bibfield  {author} {\bibinfo {author} {\bibfnamefont {M.}~\bibnamefont
			{Bibes}}, \bibinfo {author} {\bibfnamefont {K.}~\bibnamefont {Bouzehouane}},
		\bibinfo {author} {\bibfnamefont {A.}~\bibnamefont {Barth\'{e}l\'{e}my}},
		\bibinfo {author} {\bibfnamefont {M.}~\bibnamefont {Besse}}, \bibinfo
		{author} {\bibfnamefont {S.}~\bibnamefont {Fusil}}, \bibinfo {author}
		{\bibfnamefont {M.}~\bibnamefont {Bowen}}, \bibinfo {author} {\bibfnamefont
			{P.}~\bibnamefont {Seneor}}, \bibinfo {author} {\bibfnamefont
			{J.}~\bibnamefont {Carrey}}, \bibinfo {author} {\bibfnamefont
			{V.}~\bibnamefont {Cros}}, \bibinfo {author} {\bibfnamefont {A.}~\bibnamefont
			{Vaurès}}, \bibinfo {author} {\bibfnamefont {J.-P.}\ \bibnamefont
			{Contour}}, \ and\ \bibinfo {author} {\bibfnamefont {A.}~\bibnamefont
			{Fert}},\ }\href {\doibase 10.1063/1.1612902} {\bibfield  {journal} {\bibinfo
			{journal} {Appl. Phys. Lett.}\ }\textbf {\bibinfo {volume} {83}},\ \bibinfo
		{pages} {2629} (\bibinfo {year} {2003})}\BibitemShut {NoStop}%
	\bibitem [{\citenamefont {Fix}\ \emph {et~al.}(2006)\citenamefont {Fix},
		\citenamefont {Stoeffler}, \citenamefont {Henry}, \citenamefont {Colis},
		\citenamefont {Dinia}, \citenamefont {Dimopoulos}, \citenamefont {B{\"a}},\
		and\ \citenamefont {Wecker}}]{fix_diode_2006}%
	\BibitemOpen
	\bibfield  {author} {\bibinfo {author} {\bibfnamefont {T.}~\bibnamefont
			{Fix}}, \bibinfo {author} {\bibfnamefont {D.}~\bibnamefont {Stoeffler}},
		\bibinfo {author} {\bibfnamefont {Y.}~\bibnamefont {Henry}}, \bibinfo
		{author} {\bibfnamefont {S.}~\bibnamefont {Colis}}, \bibinfo {author}
		{\bibfnamefont {A.}~\bibnamefont {Dinia}}, \bibinfo {author} {\bibfnamefont
			{T.}~\bibnamefont {Dimopoulos}}, \bibinfo {author} {\bibfnamefont
			{L.}~\bibnamefont {B{\"a}}}, \ and\ \bibinfo {author} {\bibfnamefont
			{J.}~\bibnamefont {Wecker}},\ }\href {\doibase 10.1063/1.2170070} {\bibfield
		{journal} {\bibinfo  {journal} {J. Appl. Phys.}\ }\textbf {\bibinfo {volume}
			{99}},\ \bibinfo {pages} {08J107} (\bibinfo {year} {2006})}\BibitemShut
	{NoStop}%
	\bibitem [{\citenamefont {Kumar}\ \emph
		{et~al.}(2014{\natexlab{b}})\citenamefont {Kumar}, \citenamefont {Misra},
		\citenamefont {Kotnala}, \citenamefont {Gaur},\ and\ \citenamefont
		{Katiyar}}]{kumar_room_2014}%
	\BibitemOpen
	\bibfield  {author} {\bibinfo {author} {\bibfnamefont {N.}~\bibnamefont
			{Kumar}}, \bibinfo {author} {\bibfnamefont {P.}~\bibnamefont {Misra}},
		\bibinfo {author} {\bibfnamefont {R.~K.}\ \bibnamefont {Kotnala}}, \bibinfo
		{author} {\bibfnamefont {A.}~\bibnamefont {Gaur}}, \ and\ \bibinfo {author}
		{\bibfnamefont {R.~S.}\ \bibnamefont {Katiyar}},\ }\href {\doibase
		10.1088/0022-3727/47/6/065006} {\bibfield  {journal} {\bibinfo  {journal} {J.
				Phys. D: Appl. Phys.}\ }\textbf {\bibinfo {volume} {47}},\ \bibinfo {pages}
		{065006} (\bibinfo {year} {2014}{\natexlab{b}})}\BibitemShut {NoStop}%
	\bibitem [{\citenamefont {Kang}\ \emph {et~al.}(2002)\citenamefont {Kang},
		\citenamefont {Park}, \citenamefont {Lee},\ and\ \citenamefont
		{Min}}]{kang_electronic_2002}%
	\BibitemOpen
	\bibfield  {author} {\bibinfo {author} {\bibfnamefont {S.-J.}\ \bibnamefont
			{Kang}}, \bibinfo {author} {\bibfnamefont {J.~H.}\ \bibnamefont {Park}},
		\bibinfo {author} {\bibfnamefont {B.~W.}\ \bibnamefont {Lee}}, \ and\
		\bibinfo {author} {\bibfnamefont {B.~I.}\ \bibnamefont {Min}},\ }\href@noop
	{} {\bibfield  {journal} {\bibinfo  {journal} {J. Phys. Soc. Jpn.}\ }\textbf
		{\bibinfo {volume} {71}},\ \bibinfo {pages} {157} (\bibinfo {year}
		{2002})}\BibitemShut {NoStop}%
	\bibitem [{\citenamefont {Mohamed
			Musa~Saad}(2012)}]{mohamed_musa_saad_first-principles_2012}%
	\BibitemOpen
	\bibfield  {author} {\bibinfo {author} {\bibfnamefont {H.-E.}\ \bibnamefont
			{Mohamed Musa~Saad}},\ }\href {\doibase 10.1016/j.physb.2012.03.058}
	{\bibfield  {journal} {\bibinfo  {journal} {Phys. B}\ }\textbf {\bibinfo
			{volume} {407}},\ \bibinfo {pages} {2512} (\bibinfo {year}
		{2012})}\BibitemShut {NoStop}%
	\bibitem [{\citenamefont {Maignan}\ \emph {et~al.}(1999)\citenamefont
		{Maignan}, \citenamefont {Raveau}, \citenamefont {Martin},\ and\
		\citenamefont {Hervieu}}]{maignan_large_1999}%
	\BibitemOpen
	\bibfield  {author} {\bibinfo {author} {\bibfnamefont {A.}~\bibnamefont
			{Maignan}}, \bibinfo {author} {\bibfnamefont {B.}~\bibnamefont {Raveau}},
		\bibinfo {author} {\bibfnamefont {C.}~\bibnamefont {Martin}}, \ and\ \bibinfo
		{author} {\bibfnamefont {M.}~\bibnamefont {Hervieu}},\ }\href@noop {}
	{\bibfield  {journal} {\bibinfo  {journal} {J. Solid State Chem.}\ }\textbf
		{\bibinfo {volume} {144}},\ \bibinfo {pages} {224} (\bibinfo {year}
		{1999})}\BibitemShut {NoStop}%
	\bibitem [{\citenamefont {Han}\ \emph {et~al.}(2001)\citenamefont {Han},
		\citenamefont {Han}, \citenamefont {Park}, \citenamefont {Lee}, \citenamefont
		{Kim},\ and\ \citenamefont {Kim}}]{han_effect_2001}%
	\BibitemOpen
	\bibfield  {author} {\bibinfo {author} {\bibfnamefont {H.}~\bibnamefont
			{Han}}, \bibinfo {author} {\bibfnamefont {B.~J.}\ \bibnamefont {Han}},
		\bibinfo {author} {\bibfnamefont {J.~S.}\ \bibnamefont {Park}}, \bibinfo
		{author} {\bibfnamefont {B.~W.}\ \bibnamefont {Lee}}, \bibinfo {author}
		{\bibfnamefont {S.~J.}\ \bibnamefont {Kim}}, \ and\ \bibinfo {author}
		{\bibfnamefont {C.~S.}\ \bibnamefont {Kim}},\ }\href {\doibase
		10.1063/1.1362655} {\bibfield  {journal} {\bibinfo  {journal} {J. Appl.
				Phys.}\ }\textbf {\bibinfo {volume} {89}},\ \bibinfo {pages} {7687} (\bibinfo
		{year} {2001})}\BibitemShut {NoStop}%
	\bibitem [{\citenamefont {Kim}\ \emph {et~al.}(2002)\citenamefont {Kim},
		\citenamefont {Lee},\ and\ \citenamefont {Kim}}]{kim_neutron_2002}%
	\BibitemOpen
	\bibfield  {author} {\bibinfo {author} {\bibfnamefont {S.~B.}\ \bibnamefont
			{Kim}}, \bibinfo {author} {\bibfnamefont {B.~W.}\ \bibnamefont {Lee}}, \ and\
		\bibinfo {author} {\bibfnamefont {C.~S.}\ \bibnamefont {Kim}},\ }\href@noop
	{} {\bibfield  {journal} {\bibinfo  {journal} {J. Magn. Magn. Mater.}\
		}\textbf {\bibinfo {volume} {242-245}},\ \bibinfo {pages} {747} (\bibinfo
		{year} {2002})}\BibitemShut {NoStop}%
	\bibitem [{\citenamefont {Pandey}\ \emph {et~al.}(2009)\citenamefont {Pandey},
		\citenamefont {Verma}, \citenamefont {Aloysius}, \citenamefont {Bhalla},
		\citenamefont {Awana}, \citenamefont {Kishan},\ and\ \citenamefont
		{Kotnala}}]{pandey_magnetic_2009}%
	\BibitemOpen
	\bibfield  {author} {\bibinfo {author} {\bibfnamefont {V.}~\bibnamefont
			{Pandey}}, \bibinfo {author} {\bibfnamefont {V.}~\bibnamefont {Verma}},
		\bibinfo {author} {\bibfnamefont {R.~P.}\ \bibnamefont {Aloysius}}, \bibinfo
		{author} {\bibfnamefont {G.~L.}\ \bibnamefont {Bhalla}}, \bibinfo {author}
		{\bibfnamefont {V.~P.~S.}\ \bibnamefont {Awana}}, \bibinfo {author}
		{\bibfnamefont {H.}~\bibnamefont {Kishan}}, \ and\ \bibinfo {author}
		{\bibfnamefont {R.~K.}\ \bibnamefont {Kotnala}},\ }\href@noop {} {\bibfield
		{journal} {\bibinfo  {journal} {J. Magn. Magn. Mater.}\ }\textbf {\bibinfo
			{volume} {321}},\ \bibinfo {pages} {2239} (\bibinfo {year}
		{2009})}\BibitemShut {NoStop}%
\bibitem [{\citenamefont {Hemery}\ \emph {et~al.}(2007)\citenamefont {Hemery},
		\citenamefont {Williams},\ and\ \citenamefont
		{Trodahl}}]{hemery_effect_2007}%
	\BibitemOpen
	\bibfield  {author} {\bibinfo {author} {\bibfnamefont {E.}~\bibnamefont
			{Hemery}}, \bibinfo {author} {\bibfnamefont {G.~V.~M.}\ \bibnamefont
			{Williams}}, \ and\ \bibinfo {author} {\bibfnamefont {H.~J.}\ \bibnamefont
			{Trodahl}},\ }\href@noop {} {\bibfield  {journal} {\bibinfo  {journal}
			{Physica B}\ }\textbf {\bibinfo {volume} {390}},\ \bibinfo {pages} {175}
		(\bibinfo {year} {2007})}\BibitemShut {NoStop}%
	\bibitem [{\citenamefont {Yang}\ \emph {et~al.}(2003)\citenamefont {Yang},
		\citenamefont {Lee}, \citenamefont {Han}, \citenamefont {Lee},\ and\
		\citenamefont {Kim}}]{yang_enhancement_2003}%
	\BibitemOpen
	\bibfield  {author} {\bibinfo {author} {\bibfnamefont {H.~M.}\ \bibnamefont
			{Yang}}, \bibinfo {author} {\bibfnamefont {W.~Y.}\ \bibnamefont {Lee}},
		\bibinfo {author} {\bibfnamefont {H.}~\bibnamefont {Han}}, \bibinfo {author}
		{\bibfnamefont {B.~W.}\ \bibnamefont {Lee}}, \ and\ \bibinfo {author}
		{\bibfnamefont {C.~S.}\ \bibnamefont {Kim}},\ }\href@noop {} {\bibfield
		{journal} {\bibinfo  {journal} {J. Appl. Phys.}\ }\textbf {\bibinfo {volume}
			{91}},\ \bibinfo {pages} {6987} (\bibinfo {year} {2003})}\BibitemShut
	{NoStop}%
	\bibitem [{\citenamefont {McCann}\ \emph {et~al.}(2014)\citenamefont {McCann},
		\citenamefont {Williams}, \citenamefont {Hyndman}, \citenamefont {Stephen},\
		and\ \citenamefont {Kennedy}}]{mccann_large_2014}%
	\BibitemOpen
	\bibfield  {author} {\bibinfo {author} {\bibfnamefont {D.}~\bibnamefont
			{McCann}}, \bibinfo {author} {\bibfnamefont {G.}~\bibnamefont {Williams}},
		\bibinfo {author} {\bibfnamefont {A.}~\bibnamefont {Hyndman}}, \bibinfo
		{author} {\bibfnamefont {J.}~\bibnamefont {Stephen}}, \ and\ \bibinfo
		{author} {\bibfnamefont {J.}~\bibnamefont {Kennedy}},\ }\href {\doibase
		10.1016/j.physb.2013.12.008} {\bibfield  {journal} {\bibinfo  {journal}
			{Physica B}\ }\textbf {\bibinfo {volume} {436}},\ \bibinfo {pages} {126}
		(\bibinfo {year} {2014})}\BibitemShut {NoStop}%
	\bibitem [{\citenamefont {Kim}\ \emph {et~al.}(2012)\citenamefont {Kim},
		\citenamefont {Ghosh}, \citenamefont {Buvaev}, \citenamefont {Mhin},
		\citenamefont {Jones}, \citenamefont {Hebard},\ and\ \citenamefont
		{Norton}}]{kim_magnetic_2012}%
	\BibitemOpen
	\bibfield  {author} {\bibinfo {author} {\bibfnamefont {K.-W.}\ \bibnamefont
			{Kim}}, \bibinfo {author} {\bibfnamefont {S.}~\bibnamefont {Ghosh}}, \bibinfo
		{author} {\bibfnamefont {S.}~\bibnamefont {Buvaev}}, \bibinfo {author}
		{\bibfnamefont {S.}~\bibnamefont {Mhin}}, \bibinfo {author} {\bibfnamefont
			{J.~L.}\ \bibnamefont {Jones}}, \bibinfo {author} {\bibfnamefont {A.~F.}\
			\bibnamefont {Hebard}}, \ and\ \bibinfo {author} {\bibfnamefont {D.~P.}\
			\bibnamefont {Norton}},\ }\href {\doibase 10.1063/1.4761843} {\bibfield
		{journal} {\bibinfo  {journal} {J. Appl. Phys.}\ }\textbf {\bibinfo {volume}
			{112}},\ \bibinfo {pages} {083923} (\bibinfo {year} {2012})}\BibitemShut
	{NoStop}%
	\bibitem [{\citenamefont {Kim}\ \emph {et~al.}(2015)\citenamefont {Kim},
		\citenamefont {Ghosh}, \citenamefont {Buvaev}, \citenamefont {Mhin},
		\citenamefont {Jones}, \citenamefont {Hebard},\ and\ \citenamefont
		{Norton}}]{kim_effects_2015}%
	\BibitemOpen
	\bibfield  {author} {\bibinfo {author} {\bibfnamefont {K.-W.}\ \bibnamefont
			{Kim}}, \bibinfo {author} {\bibfnamefont {S.}~\bibnamefont {Ghosh}}, \bibinfo
		{author} {\bibfnamefont {S.}~\bibnamefont {Buvaev}}, \bibinfo {author}
		{\bibfnamefont {S.}~\bibnamefont {Mhin}}, \bibinfo {author} {\bibfnamefont
			{J.~L.}\ \bibnamefont {Jones}}, \bibinfo {author} {\bibfnamefont {A.~F.}\
			\bibnamefont {Hebard}}, \ and\ \bibinfo {author} {\bibfnamefont {D.~P.}\
			\bibnamefont {Norton}},\ }\href {\doibase 10.1063/1.4923354} {\bibfield
		{journal} {\bibinfo  {journal} {J. Appl. Phys.}\ }\textbf {\bibinfo {volume}
			{118}},\ \bibinfo {pages} {033903} (\bibinfo {year} {2015})}\BibitemShut
	{NoStop}%
	\bibitem [{\citenamefont {Stephen}(2013)}]{stephen_magnetic_2013}%
	\BibitemOpen
	\bibfield  {author} {\bibinfo {author} {\bibfnamefont {J.}~\bibnamefont
			{Stephen}},\ }\emph {\bibinfo {title} {Magnetic and transport properties of
			electronically polarised double perovskites and {Heusler} intermetallics}},\
	\href@noop {} {Ph.D. thesis},\ \bibinfo  {school} {Victoria University of
		Wellington} (\bibinfo {year} {2013})\BibitemShut {NoStop}%
	\bibitem [{\citenamefont {Farrell}\ \emph {et~al.}(2017)\citenamefont
		{Farrell}, \citenamefont {Hyndman}, \citenamefont {Reeves}, \citenamefont
		{Williams},\ and\ \citenamefont {Granville}}]{farrell_2017}%
	\BibitemOpen
	\bibfield  {author} {\bibinfo {author} {\bibfnamefont {I.}~\bibnamefont
			{Farrell}}, \bibinfo {author} {\bibfnamefont {A.}~\bibnamefont {Hyndman}},
		\bibinfo {author} {\bibfnamefont {R.}~\bibnamefont {Reeves}}, \bibinfo
		{author} {\bibfnamefont {G.}~\bibnamefont {Williams}}, \ and\ \bibinfo
		{author} {\bibfnamefont {S.}~\bibnamefont {Granville}},\ }\href {\doibase
		10.1016/j.tsf.2017.01.058} {\bibfield  {journal} {\bibinfo  {journal} {Thin
				Solid Films}\ }\textbf {\bibinfo {volume} {625}},\ \bibinfo {pages} {24}
		(\bibinfo {year} {2017})}\BibitemShut {NoStop}%
	\bibitem [{\citenamefont {Kim}\ \emph {et~al.}(2011)\citenamefont {Kim},
		\citenamefont {Lee}, \citenamefont {Yu}, \citenamefont {Lee},\ and\
		\citenamefont {Kang}}]{kim_sintering_2011}%
	\BibitemOpen
	\bibfield  {author} {\bibinfo {author} {\bibfnamefont {K.~S.}\ \bibnamefont
			{Kim}}, \bibinfo {author} {\bibfnamefont {M.~S.}\ \bibnamefont {Lee}},
		\bibinfo {author} {\bibfnamefont {S.~C.}\ \bibnamefont {Yu}}, \bibinfo
		{author} {\bibfnamefont {B.~W.}\ \bibnamefont {Lee}}, \ and\ \bibinfo
		{author} {\bibfnamefont {B.~S.}\ \bibnamefont {Kang}},\ }\href {\doibase
		10.1109/TMAG.2011.2160049} {\bibfield  {journal} {\bibinfo  {journal} {IEEE
				Trans. Magn.}\ }\textbf {\bibinfo {volume} {47}},\ \bibinfo {pages} {2474}
		(\bibinfo {year} {2011})}\BibitemShut {NoStop}%
	\bibitem [{\citenamefont {Ogale}\ \emph {et~al.}(1999)\citenamefont {Ogale},
		\citenamefont {Ogale}, \citenamefont {Ramesh},\ and\ \citenamefont
		{Venkatesan}}]{ogale_octahedral_1999}%
	\BibitemOpen
	\bibfield  {author} {\bibinfo {author} {\bibfnamefont {A.~S.}\ \bibnamefont
			{Ogale}}, \bibinfo {author} {\bibfnamefont {S.~B.}\ \bibnamefont {Ogale}},
		\bibinfo {author} {\bibfnamefont {R.}~\bibnamefont {Ramesh}}, \ and\ \bibinfo
		{author} {\bibfnamefont {T.}~\bibnamefont {Venkatesan}},\ }\href {\doibase
		10.1063/1.124440} {\bibfield  {journal} {\bibinfo  {journal} {Appl. Phys.
				Lett.}\ }\textbf {\bibinfo {volume} {75}},\ \bibinfo {pages} {537} (\bibinfo
		{year} {1999})}\BibitemShut {NoStop}%
	\bibitem [{\citenamefont {Balcells}\ \emph {et~al.}(2001)\citenamefont
		{Balcells}, \citenamefont {Navarro}, \citenamefont {Bibes}, \citenamefont
		{Roig}, \citenamefont {Mart{\'i}nez},\ and\ \citenamefont
		{Fontcuberta}}]{balcells_cationic_2001}%
	\BibitemOpen
	\bibfield  {author} {\bibinfo {author} {\bibfnamefont {L.}~\bibnamefont
			{Balcells}}, \bibinfo {author} {\bibfnamefont {J.}~\bibnamefont {Navarro}},
		\bibinfo {author} {\bibfnamefont {M.}~\bibnamefont {Bibes}}, \bibinfo
		{author} {\bibfnamefont {A.}~\bibnamefont {Roig}}, \bibinfo {author}
		{\bibfnamefont {B.}~\bibnamefont {Mart{\'i}nez}}, \ and\ \bibinfo {author}
		{\bibfnamefont {J.}~\bibnamefont {Fontcuberta}},\ }\href {\doibase
		10.1063/1.1346624} {\bibfield  {journal} {\bibinfo  {journal} {Appl. Phys.
				Lett.}\ }\textbf {\bibinfo {volume} {78}},\ \bibinfo {pages} {781} (\bibinfo
		{year} {2001})}\BibitemShut {NoStop}%
	\bibitem [{\citenamefont {Niebieskikwiat}\ \emph {et~al.}(2004)\citenamefont
		{Niebieskikwiat}, \citenamefont {Prado}, \citenamefont {Caneiro},\ and\
		\citenamefont {S\'{a}nchez}}]{niebieskikwiat_antisite_2004}%
	\BibitemOpen
	\bibfield  {author} {\bibinfo {author} {\bibfnamefont {D.}~\bibnamefont
			{Niebieskikwiat}}, \bibinfo {author} {\bibfnamefont {F.}~\bibnamefont
			{Prado}}, \bibinfo {author} {\bibfnamefont {A.}~\bibnamefont {Caneiro}}, \
		and\ \bibinfo {author} {\bibfnamefont {R.~D.}\ \bibnamefont {S\'{a}nchez}},\
	}\href {\doibase 10.1103/PhysRevB.70.132412} {\bibfield  {journal} {\bibinfo
			{journal} {Phys. Rev. B}\ }\textbf {\bibinfo {volume} {70}},\ \bibinfo
		{pages} {132412} (\bibinfo {year} {2004})}\BibitemShut {NoStop}%
	\bibitem [{\citenamefont {Han}\ \emph {et~al.}(2003)\citenamefont {Han},
		\citenamefont {Kim},\ and\ \citenamefont {Lee}}]{han_magnetic_2003}%
	\BibitemOpen
	\bibfield  {author} {\bibinfo {author} {\bibfnamefont {H.}~\bibnamefont
			{Han}}, \bibinfo {author} {\bibfnamefont {C.~S.}\ \bibnamefont {Kim}}, \ and\
		\bibinfo {author} {\bibfnamefont {B.~W.}\ \bibnamefont {Lee}},\ }\href
	{http://www.sciencedirect.com/science/article/pii/S0304885302009010}
	{\bibfield  {journal} {\bibinfo  {journal} {J. Magn. Magn. Mater.}\ }\textbf
		{\bibinfo {volume} {254}},\ \bibinfo {pages} {574} (\bibinfo {year}
		{2003})}\BibitemShut {NoStop}%
	\bibitem [{\citenamefont {Tovar}\ \emph {et~al.}(2002)\citenamefont {Tovar},
		\citenamefont {Causa}, \citenamefont {Butera}, \citenamefont {Navarro},
		\citenamefont {Mart\'{i}nez}, \citenamefont {Fontcuberta},\ and\
		\citenamefont {Passeggi}}]{tovar_2002}%
	\BibitemOpen
	\bibfield  {author} {\bibinfo {author} {\bibfnamefont {M.}~\bibnamefont
			{Tovar}}, \bibinfo {author} {\bibfnamefont {M.~T.}\ \bibnamefont {Causa}},
		\bibinfo {author} {\bibfnamefont {A.}~\bibnamefont {Butera}}, \bibinfo
		{author} {\bibfnamefont {J.}~\bibnamefont {Navarro}}, \bibinfo {author}
		{\bibfnamefont {B.}~\bibnamefont {Mart\'{i}nez}}, \bibinfo {author}
		{\bibfnamefont {J.}~\bibnamefont {Fontcuberta}}, \ and\ \bibinfo {author}
		{\bibfnamefont {M.~C.~G.}\ \bibnamefont {Passeggi}},\ }\href {\doibase
		10.1103/PhysRevB.66.024409} {\bibfield  {journal} {\bibinfo  {journal} {Phys.
				Rev. B}\ }\textbf {\bibinfo {volume} {66}},\ \bibinfo {pages} {024409}
		(\bibinfo {year} {2002})}\BibitemShut {NoStop}%
	\bibitem [{\citenamefont {Navarro}\ \emph {et~al.}(2003)\citenamefont
		{Navarro}, \citenamefont {Nogu{\'e}s}, \citenamefont {Mu{\~n}oz},\ and\
		\citenamefont {Fontcuberta}}]{navarro_2003}%
	\BibitemOpen
	\bibfield  {author} {\bibinfo {author} {\bibfnamefont {J.}~\bibnamefont
			{Navarro}}, \bibinfo {author} {\bibfnamefont {J.}~\bibnamefont {Nogu{\'e}s}},
		\bibinfo {author} {\bibfnamefont {J.~S.}\ \bibnamefont {Mu{\~n}oz}}, \ and\
		\bibinfo {author} {\bibfnamefont {J.}~\bibnamefont {Fontcuberta}},\ }\href
	{\doibase 10.1103/PhysRevB.67.174416} {\bibfield  {journal} {\bibinfo
			{journal} {Phys. Rev. B}\ }\textbf {\bibinfo {volume} {67}},\ \bibinfo
		{pages} {174416} (\bibinfo {year} {2003})}\BibitemShut {NoStop}%
	\bibitem [{\citenamefont {Blundell}(2001)}]{blundell_2001}%
	\BibitemOpen
	\bibfield  {author} {\bibinfo {author} {\bibfnamefont {S.}~\bibnamefont
			{Blundell}},\ }\href@noop {} {\emph {\bibinfo {title} {Magnetism in
				{Condensed} {Matter}}}}\ (\bibinfo  {publisher} {Oxford University Press},\
	\bibinfo {address} {Oxford},\ \bibinfo {year} {2001})\BibitemShut {NoStop}%
	\bibitem [{\citenamefont {Park}\ \emph {et~al.}(2001)\citenamefont {Park},
		\citenamefont {Han}, \citenamefont {Kim},\ and\ \citenamefont
		{Lee}}]{park_2001}%
	\BibitemOpen
	\bibfield  {author} {\bibinfo {author} {\bibfnamefont {J.~S.}\ \bibnamefont
			{Park}}, \bibinfo {author} {\bibfnamefont {B.~J.}\ \bibnamefont {Han}},
		\bibinfo {author} {\bibfnamefont {K.~W.}\ \bibnamefont {Kim}}, \ and\
		\bibinfo {author} {\bibfnamefont {B.~W.}\ \bibnamefont {Lee}},\ }\href@noop
	{} {\bibfield  {journal} {\bibinfo  {journal} {J. Magn. Magn. Mater.}\
		}\textbf {\bibinfo {volume} {226-230}},\ \bibinfo {pages} {741} (\bibinfo
		{year} {2001})}\BibitemShut {NoStop}%
	\bibitem [{\citenamefont {Mott}\ and\ \citenamefont {Davis}(1979)}]{mott_1979}%
	\BibitemOpen
	\bibfield  {author} {\bibinfo {author} {\bibfnamefont {N.~F.}\ \bibnamefont
			{Mott}}\ and\ \bibinfo {author} {\bibfnamefont {E.~A.}\ \bibnamefont
			{Davis}},\ }\href@noop {} {\emph {\bibinfo {title} {Electronic {Processes} in
				{Non}-{Crystalline} {Materials}}}},\ \bibinfo {edition} {2nd}\ ed.\ (\bibinfo
	{publisher} {Clarendon},\ \bibinfo {address} {Oxford},\ \bibinfo {year}
	{1979})\BibitemShut {NoStop}%
	\bibitem [{\citenamefont {Shklovskii}\ and\ \citenamefont
		{Efros}(1984)}]{shklovskii_1984}%
	\BibitemOpen
	\bibfield  {author} {\bibinfo {author} {\bibfnamefont {B.~I.}\ \bibnamefont
			{Shklovskii}}\ and\ \bibinfo {author} {\bibfnamefont {A.~L.}\ \bibnamefont
			{Efros}},\ }\href@noop {} {\emph {\bibinfo {title} {Electronic {Properties}
				of {Doped} {Semiconductors}}}}\ (\bibinfo  {publisher} {Springer},\ \bibinfo
	{address} {New York},\ \bibinfo {year} {1984})\BibitemShut {NoStop}%
	\bibitem [{\citenamefont {Sheng}(1980)}]{sheng_1980}%
	\BibitemOpen
	\bibfield  {author} {\bibinfo {author} {\bibfnamefont {P.}~\bibnamefont
			{Sheng}},\ }\href
	{http://journals.aps.org/prb/abstract/10.1103/PhysRevB.21.2180} {\bibfield
		{journal} {\bibinfo  {journal} {Phys. Rev. B}\ }\textbf {\bibinfo {volume}
			{21}},\ \bibinfo {pages} {2180} (\bibinfo {year} {1980})}\BibitemShut
	{NoStop}%
	\bibitem [{\citenamefont {Moritomo}\ \emph {et~al.}(2000)\citenamefont
		{Moritomo}, \citenamefont {Xu}, \citenamefont {Akimoto}, \citenamefont
		{Machida}, \citenamefont {Hamada}, \citenamefont {Ohoyama}, \citenamefont
		{Nishibori}, \citenamefont {Takata},\ and\ \citenamefont
		{Sakata}}]{moritomo_2000}%
	\BibitemOpen
	\bibfield  {author} {\bibinfo {author} {\bibfnamefont {Y.}~\bibnamefont
			{Moritomo}}, \bibinfo {author} {\bibfnamefont {S.}~\bibnamefont {Xu}},
		\bibinfo {author} {\bibfnamefont {T.}~\bibnamefont {Akimoto}}, \bibinfo
		{author} {\bibfnamefont {A.}~\bibnamefont {Machida}}, \bibinfo {author}
		{\bibfnamefont {N.}~\bibnamefont {Hamada}}, \bibinfo {author} {\bibfnamefont
			{K.}~\bibnamefont {Ohoyama}}, \bibinfo {author} {\bibfnamefont
			{E.}~\bibnamefont {Nishibori}}, \bibinfo {author} {\bibfnamefont
			{M.}~\bibnamefont {Takata}}, \ and\ \bibinfo {author} {\bibfnamefont
			{M.}~\bibnamefont {Sakata}},\ }\href@noop {} {\bibfield  {journal} {\bibinfo
			{journal} {Phys. Rev. B}\ }\textbf {\bibinfo {volume} {62}},\ \bibinfo
		{pages} {14224} (\bibinfo {year} {2000})}\BibitemShut {NoStop}%
	\bibitem [{\citenamefont {Saitoh}\ \emph {et~al.}(2005)\citenamefont {Saitoh},
		\citenamefont {Nakatake}, \citenamefont {Nakajima}, \citenamefont {Morimoto},
		\citenamefont {Kakizaki}, \citenamefont {Xu}, \citenamefont {Moritomo},
		\citenamefont {Hamada},\ and\ \citenamefont {Aiura}}]{saitoh_2005}%
	\BibitemOpen
	\bibfield  {author} {\bibinfo {author} {\bibfnamefont {T.}~\bibnamefont
			{Saitoh}}, \bibinfo {author} {\bibfnamefont {M.}~\bibnamefont {Nakatake}},
		\bibinfo {author} {\bibfnamefont {H.}~\bibnamefont {Nakajima}}, \bibinfo
		{author} {\bibfnamefont {O.}~\bibnamefont {Morimoto}}, \bibinfo {author}
		{\bibfnamefont {A.}~\bibnamefont {Kakizaki}}, \bibinfo {author}
		{\bibfnamefont {S.}~\bibnamefont {Xu}}, \bibinfo {author} {\bibfnamefont
			{Y.}~\bibnamefont {Moritomo}}, \bibinfo {author} {\bibfnamefont
			{N.}~\bibnamefont {Hamada}}, \ and\ \bibinfo {author} {\bibfnamefont
			{Y.}~\bibnamefont {Aiura}},\ }\href {\doibase 10.1103/PhysRevB.72.045107}
	{\bibfield  {journal} {\bibinfo  {journal} {Phys. Rev. B}\ }\textbf {\bibinfo
			{volume} {72}},\ \bibinfo {pages} {045107} (\bibinfo {year}
		{2005})}\BibitemShut {NoStop}%
	\bibitem [{\citenamefont {Popuri}\ \emph {et~al.}(2015)\citenamefont {Popuri},
		\citenamefont {Redpath}, \citenamefont {Chan}, \citenamefont {Smith},
		\citenamefont {Cespedes},\ and\ \citenamefont {Bos}}]{popuri_2015}%
	\BibitemOpen
	\bibfield  {author} {\bibinfo {author} {\bibfnamefont {S.~R.}\ \bibnamefont
			{Popuri}}, \bibinfo {author} {\bibfnamefont {D.}~\bibnamefont {Redpath}},
		\bibinfo {author} {\bibfnamefont {G.}~\bibnamefont {Chan}}, \bibinfo {author}
		{\bibfnamefont {R.~I.}\ \bibnamefont {Smith}}, \bibinfo {author}
		{\bibfnamefont {O.}~\bibnamefont {Cespedes}}, \ and\ \bibinfo {author}
		{\bibfnamefont {J.-W.~G.}\ \bibnamefont {Bos}},\ }\href@noop {} {\bibfield
		{journal} {\bibinfo  {journal} {Dalton Trans.}\ }\textbf {\bibinfo {volume}
			{44}},\ \bibinfo {pages} {10621} (\bibinfo {year} {2015})}\BibitemShut
	{NoStop}%
	\bibitem [{\citenamefont {Fisher}\ \emph {et~al.}(2003)\citenamefont {Fisher},
		\citenamefont {Chashka}, \citenamefont {Patlagan},\ and\ \citenamefont
		{Reisner}}]{fisher_2003}%
	\BibitemOpen
	\bibfield  {author} {\bibinfo {author} {\bibfnamefont {B.}~\bibnamefont
			{Fisher}}, \bibinfo {author} {\bibfnamefont {K.~B.}~\bibnamefont {Chashka}},
		\bibinfo {author} {\bibfnamefont {L.}~\bibnamefont {Patlagan}}, \ and\
		\bibinfo {author} {\bibfnamefont {G.~M.}~\bibnamefont {Reisner}},\ }\href
	{\doibase 10.1103/PhysRevB.68.134420} {\bibfield  {journal} {\bibinfo
			{journal} {Phys. Rev. B}\ }\textbf {\bibinfo {volume} {68}},\ \bibinfo
		{pages} {134420} (\bibinfo {year} {2003})}\BibitemShut {NoStop}%
	\bibitem [{\citenamefont {Fisher}\ \emph {et~al.}(2006)\citenamefont {Fisher},
		\citenamefont {Genossar}, \citenamefont {Chashka}, \citenamefont {Patlagan},\
		and\ \citenamefont {Reisner}}]{fisher_2006}%
	\BibitemOpen
	\bibfield  {author} {\bibinfo {author} {\bibfnamefont {B.}~\bibnamefont
			{Fisher}}, \bibinfo {author} {\bibfnamefont {J.}~\bibnamefont {Genossar}},
		\bibinfo {author} {\bibfnamefont {K.}~\bibnamefont {Chashka}}, \bibinfo
		{author} {\bibfnamefont {L.}~\bibnamefont {Patlagan}}, \ and\ \bibinfo
		{author} {\bibfnamefont {G.}~\bibnamefont {Reisner}},\ }\href {\doibase
		10.1016/j.ssc.2006.01.032} {\bibfield  {journal} {\bibinfo  {journal} {Solid
				State Commun.}\ }\textbf {\bibinfo {volume} {137}},\ \bibinfo {pages} {641}
		(\bibinfo {year} {2006})}\BibitemShut {NoStop}%
	\bibitem [{\citenamefont {Fisher}\ \emph {et~al.}(2014)\citenamefont {Fisher},
		\citenamefont {Genossar}, \citenamefont {Chashka}, \citenamefont {Patlagan},\
		and\ \citenamefont {Reisner}}]{fisher_2014}%
	\BibitemOpen
	\bibfield  {author} {\bibinfo {author} {\bibfnamefont {B.}~\bibnamefont
			{Fisher}}, \bibinfo {author} {\bibfnamefont {J.}~\bibnamefont {Genossar}},
		\bibinfo {author} {\bibfnamefont {K.~B.}\ \bibnamefont {Chashka}}, \bibinfo
		{author} {\bibfnamefont {L.}~\bibnamefont {Patlagan}}, \ and\ \bibinfo
		{author} {\bibfnamefont {G.~M.}\ \bibnamefont {Reisner}},\ }in\ \href
	{http://epjwoc.epj.org/articles/epjconf/abs/2014/12/epjconf_jems2014_01001/epjconf_jems2014_01001.html}
	{\emph {\bibinfo {booktitle} {{EPJ} {Web} of {Conferences}}}},\ Vol.~\bibinfo
	{volume} {75}\ (\bibinfo  {publisher} {EDP Sciences},\ \bibinfo {year}
	{2014})\ p.\ \bibinfo {pages} {01001}\BibitemShut {NoStop}%
	\bibitem [{\citenamefont {Westerburg}\ \emph
		{et~al.}(2000{\natexlab{a}})\citenamefont {Westerburg}, \citenamefont
		{Martin},\ and\ \citenamefont {Jakob}}]{westerburg_hall_2000}%
	\BibitemOpen
	\bibfield  {author} {\bibinfo {author} {\bibfnamefont {W.}~\bibnamefont
			{Westerburg}}, \bibinfo {author} {\bibfnamefont {F.}~\bibnamefont {Martin}},
		\ and\ \bibinfo {author} {\bibfnamefont {G.}~\bibnamefont {Jakob}},\ }\href
	{http://ieeexplore.ieee.org/xpls/abs_all.jsp?arnumber=5026238} {\bibfield
		{journal} {\bibinfo  {journal} {J. Appl. Phys.}\ }\textbf {\bibinfo {volume}
			{87}},\ \bibinfo {pages} {5040} (\bibinfo {year}
		{2000}{\natexlab{a}})}\BibitemShut {NoStop}%
	\bibitem [{\citenamefont {Westerburg}\ \emph
		{et~al.}(2000{\natexlab{b}})\citenamefont {Westerburg}, \citenamefont
		{Reisinger},\ and\ \citenamefont {Jakob}}]{westerburg_epitaxy_2000}%
	\BibitemOpen
	\bibfield  {author} {\bibinfo {author} {\bibfnamefont {W.}~\bibnamefont
			{Westerburg}}, \bibinfo {author} {\bibfnamefont {D.}~\bibnamefont
			{Reisinger}}, \ and\ \bibinfo {author} {\bibfnamefont {G.}~\bibnamefont
			{Jakob}},\ }\href
	{http://journals.aps.org/prb/abstract/10.1103/PhysRevB.62.R767} {\bibfield
		{journal} {\bibinfo  {journal} {Phys. Rev. B}\ }\textbf {\bibinfo {volume}
			{62}},\ \bibinfo {pages} {R767} (\bibinfo {year}
		{2000}{\natexlab{b}})}\BibitemShut {NoStop}%
	\bibitem [{\citenamefont {Nagaosa}\ \emph {et~al.}(2010)\citenamefont
		{Nagaosa}, \citenamefont {Sinova}, \citenamefont {Onoda}, \citenamefont
		{MacDonald},\ and\ \citenamefont {Ong}}]{nagaosa_2010}%
	\BibitemOpen
	\bibfield  {author} {\bibinfo {author} {\bibfnamefont {N.}~\bibnamefont
			{Nagaosa}}, \bibinfo {author} {\bibfnamefont {J.}~\bibnamefont {Sinova}},
		\bibinfo {author} {\bibfnamefont {S.}~\bibnamefont {Onoda}}, \bibinfo
		{author} {\bibfnamefont {A.~H.}\ \bibnamefont {MacDonald}}, \ and\ \bibinfo
		{author} {\bibfnamefont {N.~P.}\ \bibnamefont {Ong}},\ }\href {\doibase
		10.1103/RevModPhys.82.1539} {\bibfield  {journal} {\bibinfo  {journal} {Rev.
				Mod. Phys.}\ }\textbf {\bibinfo {volume} {82}},\ \bibinfo {pages} {1539}
		(\bibinfo {year} {2010})}\BibitemShut {NoStop}%
	\bibitem [{\citenamefont {Tian}\ \emph {et~al.}(2009)\citenamefont
		{Tian}, \citenamefont {Ye},\ and\ \citenamefont {Jin}}]{tian_2009}%
	\BibitemOpen
	\bibfield  {author} {\bibinfo {author} {\bibfnamefont {Y.}~\bibnamefont
			{Tian}}, \bibinfo {author} {\bibfnamefont {L.}~\bibnamefont {Ye}}, \ and\ \bibinfo
		{author} {\bibfnamefont {X.~F.}\ \bibnamefont {Jin}},\ }\href {\doibase
		10.1103/PhysRevLett.103.087206} {\bibfield  {journal} {\bibinfo  {journal} {Phys. Rev.
Lett.}\ }\textbf {\bibinfo {volume} {103}},\ \bibinfo {pages} {087206}
		(\bibinfo {year} {2009})}\BibitemShut {NoStop}%
	\bibitem [{\citenamefont {Aronzon}\ \emph {et~al.}(2000)\citenamefont
		{Aronzon}, \citenamefont {Rylkov}, \citenamefont {Kovalev}, \citenamefont
		{Melikhov},\citenamefont
		{Lagarkov}, \citenamefont {Sedova}, \citenamefont {Goiran}, \citenamefont
		{Negre},  \citenamefont
		{Raquet}, \ and\ \citenamefont {Leotin}}]{aronzon_2000}%
	\BibitemOpen
	\bibfield  {author} {\bibinfo {author} {\bibfnamefont {B.~A.}\ \bibnamefont
			{Aronzon}}, \bibinfo {author} {\bibfnamefont {V.~V.}\ \bibnamefont {Rylkov}},
		\bibinfo {author} {\bibfnamefont {D.~Y.}\ \bibnamefont {Kovalev}}, \bibinfo
		{author} {\bibfnamefont {E.~Z.}\ \bibnamefont {Melikhov}}, \bibinfo {author} {\bibfnamefont {A.~N.}\ \bibnamefont
			{Lagarkov}}, \bibinfo {author} {\bibfnamefont {M.~A.}\ \bibnamefont {Sedova}},
		\bibinfo {author} {\bibfnamefont {M.}~\bibnamefont {Goiran}}, \bibinfo
		{author} {\bibfnamefont {N.}~\bibnamefont {Negre}}, \bibinfo {author} {\bibfnamefont {B.}~\bibnamefont {Raquet}}, \ and\ \bibinfo
		{author} {\bibfnamefont {J.}~\bibnamefont {Leotin}},\ }\href {\doibase
		} {\bibfield  {journal} {\bibinfo  {journal} {phys. stat. sol. b}\ }\textbf {\bibinfo {volume} {218}},\ \bibinfo {pages} {169}
		(\bibinfo {year} {2000})}\BibitemShut {NoStop}%
	\bibitem [{\citenamefont {Yanagihara}\ \emph {et~al.}(2001)\citenamefont
		{Yanagihara}, \citenamefont {Salamon}, \citenamefont {Lyanda-Geller},
		\citenamefont {Xu},\ and\ \citenamefont
		{Moritomo}}]{yanagihara_magnetotransport_2001}%
	\BibitemOpen
	\bibfield  {author} {\bibinfo {author} {\bibfnamefont {H.}~\bibnamefont
			{Yanagihara}}, \bibinfo {author} {\bibfnamefont {M.~B.}\ \bibnamefont
			{Salamon}}, \bibinfo {author} {\bibfnamefont {Y.}~\bibnamefont
			{Lyanda-Geller}}, \bibinfo {author} {\bibfnamefont {S.}~\bibnamefont {Xu}}, \
		and\ \bibinfo {author} {\bibfnamefont {Y.}~\bibnamefont {Moritomo}},\ }\href
	{\doibase 10.1103/PhysRevB.64.214407} {\bibfield  {journal} {\bibinfo
			{journal} {Phys. Rev. B}\ }\textbf {\bibinfo {volume} {64}},\ \bibinfo
		{pages} {214407} (\bibinfo {year} {2001})}\BibitemShut {NoStop}%
	\bibitem [{\citenamefont {Bajpai}\ and\ \citenamefont
		{Nigam}(2007)}]{bajpai_2007}%
	\BibitemOpen
	\bibfield  {author} {\bibinfo {author} {\bibfnamefont {A.}~\bibnamefont
			{Bajpai}}\ and\ \bibinfo {author} {\bibfnamefont {A.~K.}\ \bibnamefont
			{Nigam}},\ }\href {\doibase 10.1063/1.2733621} {\bibfield  {journal}
		{\bibinfo  {journal} {J. Appl. Phys.}\ }\textbf {\bibinfo {volume} {101}},\
		\bibinfo {pages} {103911} (\bibinfo {year} {2007})}\BibitemShut {NoStop}%
	\bibitem [{\citenamefont {Cheng}\ \emph {et~al.}(2011)\citenamefont {Cheng},
		\citenamefont {Li}, \citenamefont {Wang}, \citenamefont {Luo}, \citenamefont
		{Liu},\ and\ \citenamefont {Zheng}}]{cheng_2011}%
	\BibitemOpen
	\bibfield  {author} {\bibinfo {author} {\bibfnamefont {Y.~H.}\ \bibnamefont
			{Cheng}}, \bibinfo {author} {\bibfnamefont {L.~Y.}\ \bibnamefont {Li}},
		\bibinfo {author} {\bibfnamefont {W.~H.}\ \bibnamefont {Wang}}, \bibinfo
		{author} {\bibfnamefont {X.~G.}\ \bibnamefont {Luo}}, \bibinfo {author}
		{\bibfnamefont {H.}~\bibnamefont {Liu}}, \ and\ \bibinfo {author}
		{\bibfnamefont {R.~K.}\ \bibnamefont {Zheng}},\ }\href {\doibase
		10.1063/1.3563080} {\bibfield  {journal} {\bibinfo  {journal} {J. App.
				Phys.}\ }\textbf {\bibinfo {volume} {109}},\ \bibinfo {pages} {073905}
		(\bibinfo {year} {2011})}\BibitemShut {NoStop}%
	\bibitem [{\citenamefont {Serrate}\ \emph {et~al.}(2005)\citenamefont
		{Serrate}, \citenamefont {De~Teresa}, \citenamefont {Algarabel},
		\citenamefont {Ibarra},\ and\ \citenamefont {Galibert}}]{serrate_2005}%
	\BibitemOpen
	\bibfield  {author} {\bibinfo {author} {\bibfnamefont {D.}~\bibnamefont
			{Serrate}}, \bibinfo {author} {\bibfnamefont {J.~M.}~\bibnamefont {De~Teresa}},
		\bibinfo {author} {\bibfnamefont {P.~A.}~\bibnamefont {Algarabel}}, \bibinfo
		{author} {\bibfnamefont {M.~R.}~\bibnamefont {Ibarra}}, \ and\ \bibinfo {author}
		{\bibfnamefont {J.}~\bibnamefont {Galibert}},\ }\href {\doibase
		10.1103/PhysRevB.71.104409} {\bibfield  {journal} {\bibinfo  {journal} {Phys.
				Rev. B}\ }\textbf {\bibinfo {volume} {71}},\ \bibinfo {pages} {104409}
		(\bibinfo {year} {2005})}\BibitemShut {NoStop}%
	\bibitem [{\citenamefont {Huang}\ \emph {et~al.}(2006)\citenamefont {Huang},
		\citenamefont {Karppinen}, \citenamefont {Yamauchi},\ and\ \citenamefont
		{Goodenough}}]{huang_2006}%
	\BibitemOpen
	\bibfield  {author} {\bibinfo {author} {\bibfnamefont {Y.~H.}\ \bibnamefont
			{Huang}}, \bibinfo {author} {\bibfnamefont {M.}~\bibnamefont {Karppinen}},
		\bibinfo {author} {\bibfnamefont {H.}~\bibnamefont {Yamauchi}}, \ and\
		\bibinfo {author} {\bibfnamefont {J.~B.}\ \bibnamefont {Goodenough}},\ }\href
	{\doibase 10.1103/PhysRevB.73.104408} {\bibfield  {journal} {\bibinfo
			{journal} {Phys. Rev. B}\ }\textbf {\bibinfo {volume} {73}},\ \bibinfo
		{pages} {104408} (\bibinfo {year} {2006})}\BibitemShut {NoStop}%
	\bibitem [{\citenamefont {Wang}\ \emph {et~al.}(2013)\citenamefont {Wang},
		\citenamefont {Xu}, \citenamefont {Ji}, \citenamefont {Zhang}, \citenamefont
		{Zhou}, \citenamefont {Gu}, \citenamefont {Chen}, \citenamefont {Yuan},
		\citenamefont {Yao},\ and\ \citenamefont {Chen}}]{wang_2013}%
	\BibitemOpen
	\bibfield  {author} {\bibinfo {author} {\bibfnamefont {J.-F.}\ \bibnamefont
			{Wang}}, \bibinfo {author} {\bibfnamefont {X.-J.}\ \bibnamefont {Xu}},
		\bibinfo {author} {\bibfnamefont {W.-J.}\ \bibnamefont {Ji}}, \bibinfo
		{author} {\bibfnamefont {S.-T.}\ \bibnamefont {Zhang}}, \bibinfo {author}
		{\bibfnamefont {J.}~\bibnamefont {Zhou}}, \bibinfo {author} {\bibfnamefont
			{Z.-B.}\ \bibnamefont {Gu}}, \bibinfo {author} {\bibfnamefont {Y.~B.}\
			\bibnamefont {Chen}}, \bibinfo {author} {\bibfnamefont {G.-L.}\ \bibnamefont
			{Yuan}}, \bibinfo {author} {\bibfnamefont {S.-H.}\ \bibnamefont {Yao}}, \
		and\ \bibinfo {author} {\bibfnamefont {Y.-F.}\ \bibnamefont {Chen}},\ }\href
	{\doibase 10.1039/c3ce40226f} {\bibfield  {journal} {\bibinfo  {journal}
			{CrystEngComm}\ }\textbf {\bibinfo {volume} {15}},\ \bibinfo {pages} {4601}
		(\bibinfo {year} {2013})}\BibitemShut {NoStop}%
	\bibitem [{\citenamefont {Goodenough}\ and\ \citenamefont
		{Dass}(2000)}]{goodenough_comment_2000}%
	\BibitemOpen
	\bibfield  {author} {\bibinfo {author} {\bibfnamefont {J.~B.}\ \bibnamefont
			{Goodenough}}\ and\ \bibinfo {author} {\bibfnamefont {R.~I.}\ \bibnamefont
			{Dass}},\ }\href@noop {} {\bibfield  {journal} {\bibinfo  {journal} {Int. J.
				Inorg. Mater.}\ }\textbf {\bibinfo {volume} {2}},\ \bibinfo {pages} {3}
		(\bibinfo {year} {2000})}\BibitemShut {NoStop}%
	\bibitem [{\citenamefont {Meneghini}\ \emph {et~al.}(2009)\citenamefont
		{Meneghini}, \citenamefont {Ray}, \citenamefont {Liscio}, \citenamefont
		{Bardelli}, \citenamefont {Mobilio},\ and\ \citenamefont
		{Sarma}}]{meneghini_nature_2009}%
	\BibitemOpen
	\bibfield  {author} {\bibinfo {author} {\bibfnamefont {C.}~\bibnamefont
			{Meneghini}}, \bibinfo {author} {\bibfnamefont {S.}~\bibnamefont {Ray}},
		\bibinfo {author} {\bibfnamefont {F.}~\bibnamefont {Liscio}}, \bibinfo
		{author} {\bibfnamefont {F.}~\bibnamefont {Bardelli}}, \bibinfo {author}
		{\bibfnamefont {S.}~\bibnamefont {Mobilio}}, \ and\ \bibinfo {author}
		{\bibfnamefont {D.~D.}\ \bibnamefont {Sarma}},\ }\href {\doibase
		10.1103/PhysRevLett.103.046403} {\bibfield  {journal} {\bibinfo  {journal}
			{Phys. Rev. Lett.}\ }\textbf {\bibinfo {volume} {103}},\ \bibinfo {pages}
		{046403} (\bibinfo {year} {2009})}\BibitemShut {NoStop}%
	\bibitem [{\citenamefont {McGuire}\ and\ \citenamefont
		{Potter}(1975)}]{mcguire_anisotropic_1975}%
	\BibitemOpen
	\bibfield  {author} {\bibinfo {author} {\bibfnamefont {T.~R.}\ \bibnamefont
			{McGuire}}\ and\ \bibinfo {author} {\bibfnamefont {R.~J.}\ \bibnamefont
			{Potter}},\ }\href@noop {} {\bibfield  {journal} {\bibinfo  {journal} {IEEE
				Trans. Magn.}\ }\textbf {\bibinfo {volume} {11}},\ \bibinfo {pages} {1018}
		(\bibinfo {year} {1975})}\BibitemShut {NoStop}%
	\bibitem [{\citenamefont {Kanchana}\ \emph {et~al.}(2007)\citenamefont
		{Kanchana}, \citenamefont {Vaitheeswaran}, \citenamefont {Alouani},\ and\
		\citenamefont {Delin}}]{kanchana_electronic_2007}%
	\BibitemOpen
	\bibfield  {author} {\bibinfo {author} {\bibfnamefont {V.}~\bibnamefont
			{Kanchana}}, \bibinfo {author} {\bibfnamefont {G.}~\bibnamefont
			{Vaitheeswaran}}, \bibinfo {author} {\bibfnamefont {M.}~\bibnamefont
			{Alouani}}, \ and\ \bibinfo {author} {\bibfnamefont {A.}~\bibnamefont
			{Delin}},\ }\href {\doibase 10.1103/PhysRevB.75.220404} {\bibfield  {journal}
		{\bibinfo  {journal} {Phys. Rev. B}\ }\textbf {\bibinfo {volume} {75}},\
		\bibinfo {pages} {220404} (\bibinfo {year} {2007})}\BibitemShut {NoStop}%
	\bibitem [{\citenamefont {Meetei}\ \emph {et~al.}(2013)\citenamefont {Meetei},
		\citenamefont {Erten}, \citenamefont {Mukherjee}, \citenamefont {Randeria},
		\citenamefont {Trivedi},\ and\ \citenamefont
		{Woodward}}]{meetei_theory_2013}%
	\BibitemOpen
	\bibfield  {author} {\bibinfo {author} {\bibfnamefont {O.~N.}\ \bibnamefont
			{Meetei}}, \bibinfo {author} {\bibfnamefont {O.}~\bibnamefont {Erten}},
		\bibinfo {author} {\bibfnamefont {A.}~\bibnamefont {Mukherjee}}, \bibinfo
		{author} {\bibfnamefont {M.}~\bibnamefont {Randeria}}, \bibinfo {author}
		{\bibfnamefont {N.}~\bibnamefont {Trivedi}}, \ and\ \bibinfo {author}
		{\bibfnamefont {P.}~\bibnamefont {Woodward}},\ }\href {\doibase
		10.1103/PhysRevB.87.165104} {\bibfield  {journal} {\bibinfo  {journal} {Phys.
				Rev. B}\ }\textbf {\bibinfo {volume} {87}},\ \bibinfo {pages} {165104}
		(\bibinfo {year} {2013})}\BibitemShut {NoStop}%
	\bibitem [{\citenamefont {Kokado}\ \emph {et~al.}(2012)\citenamefont {Kokado},
		\citenamefont {Tsunoda}, \citenamefont {Harigaya},\ and\ \citenamefont
		{Sakuma}}]{kokado_anisotropic_2012}%
	\BibitemOpen
	\bibfield  {author} {\bibinfo {author} {\bibfnamefont {S.}~\bibnamefont
			{Kokado}}, \bibinfo {author} {\bibfnamefont {M.}~\bibnamefont {Tsunoda}},
		\bibinfo {author} {\bibfnamefont {K.}~\bibnamefont {Harigaya}}, \ and\
		\bibinfo {author} {\bibfnamefont {A.}~\bibnamefont {Sakuma}},\ }\href
	{\doibase 10.1143/JPSJ.81.024705} {\bibfield  {journal} {\bibinfo  {journal}
			{J. Phys. Soc. Jpn.}\ }\textbf {\bibinfo {volume} {81}},\ \bibinfo {pages}
		{024705} (\bibinfo {year} {2012})}\BibitemShut {NoStop}%
	\bibitem [{\citenamefont {Kokado}\ and\ \citenamefont
		{Tsunoda}(2013)}]{kokado_anisotropic_2013}%
	\BibitemOpen
	\bibfield  {author} {\bibinfo {author} {\bibfnamefont {S.}~\bibnamefont
			{Kokado}}\ and\ \bibinfo {author} {\bibfnamefont {M.}~\bibnamefont
			{Tsunoda}},\ }\href {\doibase 10.4028/www.scientific.net/AMR.750-752.978}
	{\bibfield  {journal} {\bibinfo  {journal} {Adv. Mater. Res.}\ }\textbf
		{\bibinfo {volume} {750-752}},\ \bibinfo {pages} {978} (\bibinfo {year}
		{2013})}\BibitemShut {NoStop}%
	\bibitem [{\citenamefont {Yang}\ \emph {et~al.}(2012)\citenamefont {Yang},
		\citenamefont {Sakuraba}, \citenamefont {Kokado}, \citenamefont {Kota},
		\citenamefont {Sakuma},\ and\ \citenamefont
		{Takanashi}}]{yang_anisotropic_2012}%
	\BibitemOpen
	\bibfield  {author} {\bibinfo {author} {\bibfnamefont {F.~J.}\ \bibnamefont
			{Yang}}, \bibinfo {author} {\bibfnamefont {Y.}~\bibnamefont {Sakuraba}},
		\bibinfo {author} {\bibfnamefont {S.}~\bibnamefont {Kokado}}, \bibinfo
		{author} {\bibfnamefont {Y.}~\bibnamefont {Kota}}, \bibinfo {author}
		{\bibfnamefont {A.}~\bibnamefont {Sakuma}}, \ and\ \bibinfo {author}
		{\bibfnamefont {K.}~\bibnamefont {Takanashi}},\ }\href {\doibase
		10.1103/PhysRevB.86.020409} {\bibfield  {journal} {\bibinfo  {journal} {Phys.
				Rev. B}\ }\textbf {\bibinfo {volume} {86}},\ \bibinfo {pages} {020409}
		(\bibinfo {year} {2012})}\BibitemShut {NoStop}%
	\bibitem [{\citenamefont {Du}\ \emph {et~al.}(2013)\citenamefont {Du},
		\citenamefont {Xu}, \citenamefont {Liu}, \citenamefont {Li}, \citenamefont
		{Zhang}, \citenamefont {Yu}, \citenamefont {Wang},\ and\ \citenamefont
		{Wu}}]{du_half-metallicity_2013}%
	\BibitemOpen
	\bibfield  {author} {\bibinfo {author} {\bibfnamefont {Y.}~\bibnamefont
			{Du}}, \bibinfo {author} {\bibfnamefont {G.~Z.}\ \bibnamefont {Xu}}, \bibinfo
		{author} {\bibfnamefont {E.~K.}\ \bibnamefont {Liu}}, \bibinfo {author}
		{\bibfnamefont {G.~J.}\ \bibnamefont {Li}}, \bibinfo {author} {\bibfnamefont
			{H.~G.}\ \bibnamefont {Zhang}}, \bibinfo {author} {\bibfnamefont {S.~Y.}\
			\bibnamefont {Yu}}, \bibinfo {author} {\bibfnamefont {W.~H.}\ \bibnamefont
			{Wang}}, \ and\ \bibinfo {author} {\bibfnamefont {G.~H.}\ \bibnamefont
			{Wu}},\ }\href@noop {} {\bibfield  {journal} {\bibinfo  {journal} {J. Magn.
				Magn. Mater.}\ }\textbf {\bibinfo {volume} {335}},\ \bibinfo {pages} {101}
		(\bibinfo {year} {2013})}\BibitemShut {NoStop}%
	\bibitem [{\citenamefont {Sakuraba}\ \emph {et~al.}(2014)\citenamefont
		{Sakuraba}, \citenamefont {Kokado}, \citenamefont {Hirayama}, \citenamefont
		{Furubayashi}, \citenamefont {Sukegawa}, \citenamefont {Li}, \citenamefont
		{Takahashi},\ and\ \citenamefont {Hono}}]{sakuraba_quantitative_2014}%
	\BibitemOpen
	\bibfield  {author} {\bibinfo {author} {\bibfnamefont {Y.}~\bibnamefont
			{Sakuraba}}, \bibinfo {author} {\bibfnamefont {S.}~\bibnamefont {Kokado}},
		\bibinfo {author} {\bibfnamefont {Y.}~\bibnamefont {Hirayama}}, \bibinfo
		{author} {\bibfnamefont {T.}~\bibnamefont {Furubayashi}}, \bibinfo {author}
		{\bibfnamefont {H.}~\bibnamefont {Sukegawa}}, \bibinfo {author}
		{\bibfnamefont {S.}~\bibnamefont {Li}}, \bibinfo {author} {\bibfnamefont
			{Y.~K.}\ \bibnamefont {Takahashi}}, \ and\ \bibinfo {author} {\bibfnamefont
			{K.}~\bibnamefont {Hono}},\ }\href {\doibase 10.1063/1.4874851} {\bibfield
		{journal} {\bibinfo  {journal} {Appl. Phys. Lett.}\ }\textbf {\bibinfo
			{volume} {104}},\ \bibinfo {pages} {172407} (\bibinfo {year}
		{2014})}\BibitemShut {NoStop}%
	\bibitem [{\citenamefont {Bosu}\ \emph {et~al.}(2016)\citenamefont {Bosu},
		\citenamefont {Sakuraba}, \citenamefont {Sasaki}, \citenamefont {Li},\ and\
		\citenamefont {Hono}}]{bosu_enhancement_2016}%
	\BibitemOpen
	\bibfield  {author} {\bibinfo {author} {\bibfnamefont {S.}~\bibnamefont
			{Bosu}}, \bibinfo {author} {\bibfnamefont {Y.}~\bibnamefont {Sakuraba}},
		\bibinfo {author} {\bibfnamefont {T.}~\bibnamefont {Sasaki}}, \bibinfo
		{author} {\bibfnamefont {S.}~\bibnamefont {Li}}, \ and\ \bibinfo {author}
		{\bibfnamefont {K.}~\bibnamefont {Hono}},\ }\href {\doibase
		10.1016/j.scriptamat.2015.08.003} {\bibfield  {journal} {\bibinfo  {journal}
			{Scr. Mater.}\ }\textbf {\bibinfo {volume} {110}},\ \bibinfo {pages} {70}
		(\bibinfo {year} {2016})}\BibitemShut {NoStop}%
	\bibitem [{\citenamefont {Li}\ \emph {et~al.}(2010{\natexlab{a}})\citenamefont
		{Li}, \citenamefont {Jiang},\ and\ \citenamefont {Bai}}]{li_fourfold_2010}%
	\BibitemOpen
	\bibfield  {author} {\bibinfo {author} {\bibfnamefont {P.}~\bibnamefont
			{Li}}, \bibinfo {author} {\bibfnamefont {E.~Y.}\ \bibnamefont {Jiang}}, \
		and\ \bibinfo {author} {\bibfnamefont {H.~L.}\ \bibnamefont {Bai}},\ }\href
	{\doibase 10.1063/1.3334722} {\bibfield  {journal} {\bibinfo  {journal}
			{Appl. Phys. Lett.}\ }\textbf {\bibinfo {volume} {96}},\ \bibinfo {pages}
		{092502} (\bibinfo {year} {2010}{\natexlab{a}})}\BibitemShut {NoStop}%
	\bibitem [{\citenamefont {Li}\ \emph {et~al.}(2010{\natexlab{b}})\citenamefont
		{Li}, \citenamefont {Jin}, \citenamefont {Jiang},\ and\ \citenamefont
		{Bai}}]{li_origin_2010}%
	\BibitemOpen
	\bibfield  {author} {\bibinfo {author} {\bibfnamefont {P.}~\bibnamefont
			{Li}}, \bibinfo {author} {\bibfnamefont {C.}~\bibnamefont {Jin}}, \bibinfo
		{author} {\bibfnamefont {E.~Y.}\ \bibnamefont {Jiang}}, \ and\ \bibinfo
		{author} {\bibfnamefont {H.~L.}\ \bibnamefont {Bai}},\ }\href {\doibase
		10.1063/1.3499696} {\bibfield  {journal} {\bibinfo  {journal} {J. Appl.
				Phys.}\ }\textbf {\bibinfo {volume} {108}},\ \bibinfo {pages} {093921}
		(\bibinfo {year} {2010}{\natexlab{b}})}\BibitemShut {NoStop}%
	\bibitem [{\citenamefont {Jin}\ \emph {et~al.}(2012)\citenamefont {Jin},
		\citenamefont {Li}, \citenamefont {Mi},\ and\ \citenamefont
		{Bai}}]{jin_magnetocrystalline_2012}%
	\BibitemOpen
	\bibfield  {author} {\bibinfo {author} {\bibfnamefont {C.}~\bibnamefont
			{Jin}}, \bibinfo {author} {\bibfnamefont {P.}~\bibnamefont {Li}}, \bibinfo
		{author} {\bibfnamefont {W.~B.}\ \bibnamefont {Mi}}, \ and\ \bibinfo {author}
		{\bibfnamefont {H.~L.}\ \bibnamefont {Bai}},\ }\href@noop {} {\bibfield
		{journal} {\bibinfo  {journal} {Europhys. Letters}\ }\textbf {\bibinfo
			{volume} {100}},\ \bibinfo {pages} {27006} (\bibinfo {year}
		{2012})}\BibitemShut {NoStop}%
\end{thebibliography}%


\end{document}